%% file: ANA-EXOT-2019-08-PAPER.tex
\documentclass[cernpreprint, atlasdraft=false, texlive=2023, UKenglish,orcidlogo, texmf, orcidlogo]{atlasdoc}
 
\usepackage[subcaption=true]{atlaspackage}
 
\usepackage{atlasbiblatex}
 
\usepackage{atlasphysics}

\graphicspath{{logos/}{figures/}}
 
\usepackage{ANA-EXOT-2019-08-PAPER-defs}

\AtlasTitle{Search for pair-produced vector-like top and bottom partners in events with
large missing transverse momentum in $pp$ collisions with the ATLAS detector}

\AtlasAbstract{
A search for pair-produced vector-like quarks using events with exactly one lepton ($e$ or $\mu$), at least four jets including at least one $b$-tagged jet, and large missing transverse momentum is presented. Data from proton--proton collisions at a centre-of-mass energy of $\sqrt{s}=$\SI{13}{\TeV}, recorded by the ATLAS detector at the LHC from 2015 to 2018 and corresponding to an integrated luminosity of \SI{139}{\mbox{fb\(^{-1}\)}}, are analysed. Vector-like partners $T$ and $B$ of the top and bottom quarks are considered, as is a vector-like $X$ with charge $+5/3$, assuming their decay into a $W$, $Z$, or Higgs boson and a third-generation quark. No significant deviations from the Standard Model expectation are observed.
Upper limits on the production cross-section of $T$ and $B$ quark pairs as a function of their mass are derived for
various decay branching ratio scenarios.
The strongest lower limits on the masses are \SI{1.59}{\TeV} assuming mass-degenerate vector-like quarks and
branching ratios corresponding to the weak-isospin doublet model, and \SI{1.47}{\TeV} (\SI{1.46}{\TeV})
for exclusive $T \rightarrow Zt$ ($B/X \rightarrow Wt$) decays.
In addition, lower limits on the $T$ and $B$ quark masses are derived for all possible branching ratios.
}

\AtlasRefCode{EXOT-2019-08}
 
\PreprintIdNumber{CERN-EP-2022-201}

\AtlasJournal{EPJ\ C}
\AtlasJournalRef{Eur. Phys. J. C 83 (2023) 719}
\AtlasDOI{10.1140/epjc/s10052-023-11790-7}

\hypersetup{pdftitle={ATLAS document},pdfauthor={The ATLAS Collaboration}}
 
\begin{document}
 
\maketitle
 
\tableofcontents
 
\section{Introduction}
\label{sec:intro}


The fine-tuning or naturalness problem~\cite{Susskind:1978ms} in particle physics arises from
loop corrections to the Higgs boson mass that are quadratically divergent.
If the Standard Model were complete, those would lead to corrections to the Higgs mass of the order
of the Planck scale, $10^{19}$~\GeV, and a finely tuned bare Higgs mass of a similar value is needed
to arrive at the measured mass of about \SI{125}{\GeV}~\cite{HIGG-2014-14}.
 
Vector-like quarks (VLQs)~\cite{delAguila:1982fs,Aguilar-Saavedra:2009xmz,Okada:2012gy,Aguilar-Saavedra:2013qpa} could
dampen the unnaturally large quadratic corrections to the Higgs boson mass by contributing significantly to loop corrections.
They are hypothetical spin-1/2 coloured particles
whose left-handed and right-handed states have the same electroweak coupling.
They appear in a number of theories beyond the Standard Model (SM) of particle physics,
mainly in the `Little Higgs'~\cite{Arkani-Hamed:2001nha,Arkani-Hamed:2002ikv,Schmaltz:2005ky}
and `Composite Higgs'~\cite{Agashe:2004rs,Kaplan:1983sm} classes of models.

At the Large Hadron Collider (LHC), VLQs could be produced singly via electroweak interactions or in pairs mainly via the strong
interaction. While the cross-section for the latter depends only on the VLQ mass, the former has an additional dependence on the
unknown coupling strength between the electroweak bosons and the VLQ.
VLQs are expected to preferentially couple to third-generation quarks~\cite{delAguila:1982fs,Aguilar-Saavedra:2013wba}.
Therefore the up-type VLQ $T$ with charge $+2/3$ is in the following assumed to only have
the three possible decay modes $T\rightarrow Zt$, $T\rightarrow Ht$, and $T\rightarrow Wb$.
Similarly, the down-type VLQ $B$ with charge $-1/3$ can decay into $Zb$, $Hb$, or $Wt$.
Vector-like $X$ quarks with charge $+5/3$ also appear in multiplets with $T$ partners~\cite{Contino:2006qr,Aguilar-Saavedra:2009xmz}
and decay via $X \rightarrow Wt$ only.

\begin{figure}
\begin{center}
\begin{subfigure}[b]{0.5\textwidth}
\centering
\includegraphics[width=0.9\linewidth]{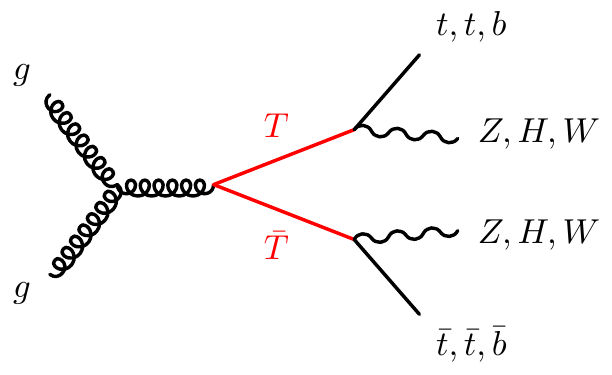}
\caption{}
\label{fig:VLT_prod}
\end{subfigure}\begin{subfigure}[b]{0.5\textwidth}
\centering
\includegraphics[width=0.9\linewidth]{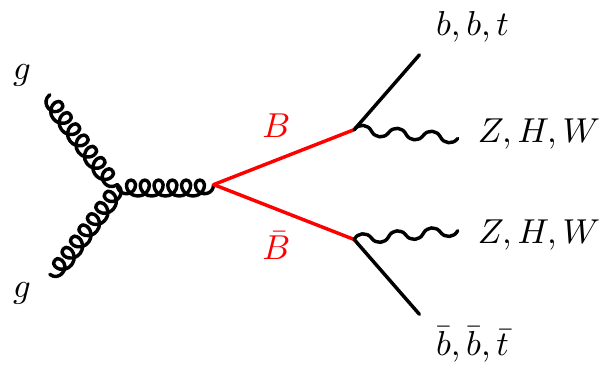}
\caption{}
\label{fig:VLB_prod}
\end{subfigure}
\caption{
Representative Feynman diagrams for \subref{fig:VLT_prod} $T\bar{T}$ and \subref{fig:VLB_prod} $B\bar{B}$ production and decay. In the analysis, no distinction is made between particles and antiparticles, leading to sensitivity
to the $X\bar{X} \rightarrow W t W  t$ final state as well.
}
\label{fig:feynman-diagrams}
\end{center}
\end{figure}

This analysis investigates all possible decay modes and combinations of branching ratios for the pair-produced vector-like $T$ (VLT) quark and $B$ (VLB) quark,
shown in Figure~\ref{fig:feynman-diagrams}.
However, it is most sensitive to the $T\rightarrow Zt$ and $B\rightarrow Wt$ decays.
Since the analysis does not distinguish between particles and
antiparticles, the limits for $B\rightarrow Wt$ also apply to the vector-like $X$ quark, given that it exclusively decays into $Wt$.
Particular combinations of branching ratios correspond to the weak-isospin singlet and doublet models.
For $T$ and $B$ quarks, the branching ratio for each decay mode depends on the VLQ mass and weak-isospin quantum numbers~\cite{Aguilar-Saavedra:2009xmz}.
The branching ratios given in the following are for VLQ masses above \SI{800}{\GeV}, where they are approximately independent of the VLQ mass.
For a singlet $T$, all decay modes have sizeable branching ratios ($\mathcal{B}(Zt,Ht,Wb) \approx (0.25,0.25,0.5)$),
whereas if $T$ is in either an $(X,T)$ doublet or a $(T,B)$ doublet, it decays only into $Zt$ or $Ht$ with equal branching ratios
as long as the generalised Cabibbo--Kobayashi--Maskawa (CKM) matrix elements fulfil $|V_{Tb}| \ll |V_{Bt}|$~\cite{Aguilar-Saavedra:2009xmz}.
Similarly, for a singlet $B$ the branching ratios of all decay modes are sizeable ($\mathcal{B}(Zb,Hb,Wt) \approx (0.25,0.25,0.5)$),
while for the $(T,B)$ doublet scenario with $|V_{Tb}| \ll |V_{Bt}|$ the $B\rightarrow Wt$ decay is the only possibility.

The results are based on the full \RunTwo (2015--2018) dataset collected by the ATLAS experiment
in $\sqrt{s}=$\SI{13}{\TeV} proton--proton collisions at the LHC, corresponding to an integrated luminosity of \SI{139}{\ifb}.
Several searches for pair-produced VLQs targeting various final states have already been performed by
ATLAS~\cite{EXOT-2016-13,EXOT-2016-14,EXOT-2016-15,EXOT-2016-16,EXOT-2016-35,EXOT-2017-14,EXOT-2017-34,ATLAS:2022hnn}
and CMS~\cite{CMS-B2G-17-011,CMS-B2G-17-012,CMS-B2G-17-003,CMS-B2G-19-005}.
Those results are based on a subset of the \RunTwo data of about \SI{36}{\ifb}, except for Refs.~\cite{ATLAS:2022hnn,CMS-B2G-19-005}.
A combination of the ATLAS results using the \SI{36}{\ifb} dataset is also available~\cite{EXOT-2017-17},
excluding $T$ ($B$) masses below \SI{1.31}{\TeV} (\SI{1.03}{\TeV}) for any combination of the three decay modes per VLQ discussed above.

The analysis is based on a final-state signature with high missing transverse momentum \MET,
one lepton~$\ell$ ($e$ or $\mu$), and at least four jets including a $b$-tagged jet.
The previous ATLAS search in this final state is based on a subset
of the \RunTwo data~\cite{EXOT-2016-15},
yielding lower limits on the $T$ quark mass of \SI{1.16}{\TeV} for $\mathcal{B}(T\rightarrow Zt) = 100$\%
and \SI{0.87}{\TeV} (\SI{1.05}{\TeV}) for the $T$ in the singlet (doublet) model.
Here the analysis is extended mainly by also investigating vector-like $B$ quarks
and by using neural networks (NNs) trained at several branching ratios
in order to separate signal from background, instead of using a
cut-and-count analysis
with a single signal region (SR).
The training  is done separately for $T$ and $B$ in a common training region using simulated events.
The SRs are each defined by a subset of the training region passing a selection on the corresponding NN output.
Control regions (CRs) are defined so as to be enriched in the various background processes.
They are orthogonal to the training region, and thus to the
SRs, and orthogonal to each other. The statistical interpretation
is based on a simultaneous fit to the CRs and SR for $T$ or $B$, in which the normalisations for
\ttbar, \Wjets and single-top-quark backgrounds and a possible
signal contribution are determined.


\section{ATLAS detector}
\label{sec:detector}

The ATLAS experiment~\cite{PERF-2007-01} at the LHC is a multipurpose particle detector with a forward--backward symmetric cylindrical geometry and a near \(4\pi\) coverage in solid angle.\footnote{\AtlasCoordFootnote}
It consists of an inner tracking detector (ID) surrounded by a thin superconducting solenoid providing a \SI{2}{\tesla} axial magnetic field, electromagnetic and hadron calorimeters, and a muon spectrometer.
The inner tracking detector covers the pseudorapidity range \(|\eta| < 2.5\).
It consists of silicon pixel, silicon microstrip, and transition radiation tracking detectors.
Lead/liquid-argon (LAr) sampling calorimeters provide electromagnetic (EM) energy measurements with high granularity.
A steel/scintillator-tile hadron calorimeter covers the central pseudorapidity range (\(|\eta| < 1.7\)).
The endcap and forward regions are instrumented with LAr calorimeters for both the EM and hadronic energy measurements up to \(|\eta| = 4.9\).
The muon spectrometer surrounds the calorimeters and is based on three large superconducting air-core toroidal magnets with eight coils each.
The field integral of the toroids ranges between \num{2.0} and \SI{6.0}{\tesla\metre} across most of the detector.
The muon spectrometer includes a system of precision tracking chambers and fast detectors for triggering.
A two-level trigger system is used to select events.
The first-level trigger is implemented in hardware and uses a subset of the detector information to accept events at a rate below \SI{100}{\kHz}.
This is followed by a software-based high-level trigger (HLT) that reduces the accepted event rate to \SI{1}{\kHz} on average depending on the data-taking conditions.
An extensive software suite~\cite{ATL-SOFT-PUB-2021-001} is used in data simulation, in the reconstruction and analysis of real and simulated data, in detector operations, and in the trigger and data acquisition systems of the experiment.


\section{Data and simulated event samples}
\label{sec:data_simulation}

The analysis uses data from proton--proton ($pp$) collisions at $\sqrt{s} = \SI{13}{\TeV}$ recorded with the ATLAS detector at the LHC in the years 2015 to 2018. The dataset, collected during stable beam conditions and with all detector subsystems operational~\cite{DAPR-2018-01}, corresponds to an integrated luminosity of \SI{139}{\ifb} with an uncertainty of \SI{1.7}{\percent}~\cite{ATLAS-CONF-2019-021}. At the high luminosities reached at the LHC, events are affected by additional inelastic $pp$ collisions in the same or neighbouring bunch crossings, referred to as pile-up. The average number of interactions per bunch crossing was 33.7.
Events were selected online during data-taking by \MET triggers~\cite{TRIG-2019-01} with an \MET threshold
of \SI{70}{\GeV} in the HLT in 2015 and a threshold rising
from \SI{90}{\GeV} to \SI{110}{\GeV} during the later years.

Monte Carlo (MC) simulated events are used for the modelling of the background processes and the VLQ signals. Details of the simulated nominal samples, including the matrix-element (ME) generator and the parton distribution function (PDF) set, the parton shower (PS) and hadronisation model, and the set of tuned parameters (tune), are summarised in Table~\ref{tab:samples:generators}.
 
\begin{table}
\centering
\caption{List of ME generator, PDF set, PS model, and tune for the signal and different background processes.}
\label{tab:samples:generators}
\resizebox{\linewidth}{!}{\begin{tabular}{lllll}
\toprule
Process           & ME generator        & PDF set    & PS and hadronisation & Tune  \\ \midrule
VLQ signal        & \PROTOS~v2.2        & \nnpdflo   & \PYTHIA[8.186]       & A14      \\
\ttbar/\singletop & \POWHEGBOX~v2       & \nnpdfnlo  & \PYTHIA[8.230/8.235]       & A14      \\
\Vjets            & \SHERPA~2.2.1       & \nnpdfnnlo & \SHERPA              & internal \\
Diboson           & \SHERPA~2.2.1--2.2.2 & \nnpdfnnlo & \SHERPA              & internal \\
\ttV              & \MGNLO[2.3.3]       & \nnpdfnlo  & \PYTHIA[8.210]       & A14      \\
\tWZnunu              & \MGNLO[2.6.7]       & \nnpdfnlo  & \PYTHIA[8.244]       & A14      \\
\ttH              & \POWHEGBOX~v2       & \nnpdfnlo  & \PYTHIA[8.230]       & A14      \\ \bottomrule
\end{tabular}}
\end{table}
 
The generated events were processed through a simulation~\cite{SOFT-2010-01} of the ATLAS detector geometry and response using \GEANT~\cite{Agostinelli:2002hh}. A faster simulation, which employed a parameterisation of the calorimeter response, was used in some cases to estimate systematic uncertainties. In these cases, the systematically varied samples were compared with versions of the nominal samples
that were also processed through the fast simulation.
In order to model \pileup effects,
minimum-bias $pp$ interactions were generated
with \PYTHIA[8.186]~\cite{Sjostrand:2007gs} using the A3~\cite{ATL-PHYS-PUB-2016-017} set of tuned parameters and overlaid on the simulated hard-scatter events. The resulting events were weighted to match the \pileup profile of the recorded data.
 
Finally, the simulated events were reconstructed using the same software as the collision data.
Corrections were applied to the simulated events in order to match object identification efficiencies, energy scales and resolution to those determined from data in auxiliary measurements.

Signal samples for the pair-production of vector-like $T$ and $B$ quarks were generated at leading order (LO) with  \PROTOS[v2.2]~\cite{protos} using the \NNPDF[2.3lo] PDF~\cite{Ball:2012cx} set, interfaced with \PYTHIA[8.186] to model the parton shower, hadronisation, and underlying event.
Using the narrow-width approximation, the samples were produced for masses from \SI{800}{\GeV} up to \SI{2}{\TeV}, with a mass spacing of \SI{100}{\GeV} from \SI{1}{\TeV} to \SI{1.8}{\TeV}.
The chirality-dependent couplings of the VLQs were set to those in the weak-isospin singlet model, but with equal branching ratios into the three decay modes ($Zt$, $Ht$, $Wb$) for the vector-like $T$ quark and ($Zb$, $Hb$, $Wt$) for the vector-like $B$ quark.
Dedicated signal samples in the doublet model were produced for the \SI{1.2}{\TeV}  mass point in order to test for potential kinematic biases from the assumed singlet couplings. For the $T$ quark, this choice is conservative because the acceptance is higher in the doublet case, while for the $B$ quark the acceptances are similar for singlet and doublet couplings.
In order to obtain the desired branching ratios, an event-by-event reweighting based on generator information is performed. The signal sample cross-sections were calculated with \TOPpp[2.0]~\cite{Czakon:2011xx} at next-to-next-to-leading order (NNLO)
in QCD including the resummation of next-to-next-to-leading logarithmic (NNLL) soft-gluon terms.

The production of \ttbar\ events was modelled using the \POWHEGBOX[v2]~\cite{Frixione:2007nw,Nason:2004rx,Frixione:2007vw,Alioli:2010xd} generator at next-to-leading order (\NLO) with the \NNPDF[3.0nlo] set~\cite{NNPDF:2014otw} of PDFs and the \hdamp parameter\footnote{The \hdamp parameter is a resummation damping factor and one of the parameters that controls the matching of \POWHEG matrix elements to the PS and thus regulates the high-\pt radiation against which the \ttbar system recoils.} set to $1.5\,m_{t}$~\cite{ATL-PHYS-PUB-2016-020}, with $m_{t}= \SI{172.5}{\GeV}$. The events were interfaced to \PYTHIA[8.230].
The cross-section was corrected to the theory prediction at \NNLO including \NNLL soft-gluon terms calculated using
\TOPpp[2.0].

Samples of single-top-quark events were produced with the \POWHEGBOX[v2] generator at \NLO in QCD using the \NNPDF[3.0nlo] set of PDFs with the five-flavour scheme for $tW$ production and $s$-channel single-top production, and the four-flavour scheme for $t$-channel single-top events. The $tW$ sample was modelled using the diagram removal scheme~\cite{Frixione:2008yi} to remove interference and overlap with \ttbar production.
The events were interfaced with either \PYTHIA[8.230] or \PYTHIA[8.235].
The samples were normalised to their \NLO QCD cross-sections~\cite{Aliev:2010zk,Kant:2014oha},
with additional \NNLL soft-gluon terms for $tW$ production~\cite{Kidonakis:2010ux,Kidonakis:2013zqa}.
 
The production of \Vjets ($V=W,Z$) was simulated with the \SHERPA[2.2.1] generator using \NLO-accurate matrix elements for up to two partons, and LO matrix elements for up to four partons, calculated with the Comix~\cite{Gleisberg:2008fv} and \OPENLOOPS~\cite{Buccioni:2019sur,Cascioli:2011va,Denner:2016kdg} libraries.
They were matched with the \SHERPA PS~\cite{Schumann:2007mg} using the \MEPSatNLO prescription~\cite{Hoeche:2011fd,Hoeche:2012yf,Catani:2001cc,Hoeche:2009rj} and the set of tuned parameters developed by the \SHERPA authors.
The \NNPDF[3.0nnlo] set of PDFs was used and the samples were normalised to a \NNLO prediction~\cite{Anastasiou:2003ds}.
 
Samples of diboson final states ($VV$) were simulated with the \SHERPA[2.2.1] or 2.2.2 generator, depending on the process, including off-shell effects and Higgs boson contributions where appropriate.
Fully leptonic final states and semileptonic final states, where one boson decays leptonically and the other one hadronically, were generated using matrix elements at \NLO accuracy in QCD for up to one additional parton and at \LO accuracy for up to three additional parton emissions.
The matching of \NLO matrix elements to the PS and the merging of different jet multiplicities was done in the same way as for \Vjets production.
The \NNPDF[3.0nnlo] set of PDFs was used, along with the \SHERPA-internal tune. The diboson event samples were normalised to the total
cross-section calculated by \SHERPA at \NLO in QCD.
 
The production of \ttW and \ttZ events was modelled using the \MGNLO[2.3.3]~\cite{Alwall:2014hca} generator at \NLO with the \NNPDF[3.0nlo] PDF set. The events were interfaced to \PYTHIA[8.210].
Similarly, the production of \tWZ events was modelled using \MGNLO[2.6.8] at \NLO with the \NNPDF[3.0nlo] PDF set, interfaced to  \PYTHIA[8.244]. The diagram removal scheme was employed to handle the interference between \tWZ and \ttZ, and was applied to the \tWZ sample. 
Samples for the production of \ttH events were generated using the \POWHEGBOX[v2] generator at \NLO with the \NNPDF[3.0nlo] PDF set, interfaced to \PYTHIA[8.230].
The generated samples for \ttW, \ttZ, \tWZ, and \ttH production were normalised to \NLO cross-section predictions calculated by \MGNLO.
 
All simulated samples, except those produced with the \SHERPA~\cite{Bothmann:2019yzt} event generator, utilised the \EVTGEN~\cite{Lange:2001uf} program to model the decay of heavy-flavour hadrons. While \EVTGEN[1.2.0] was used for the VLQ signal samples and the \ttV samples, \EVTGEN[1.6.0] was used in all other cases. For all nominal samples where \PYTHIA[8]~\cite{Sjostrand:2014zea} was utilised for the showering and hadronisation, \PYTHIA was used with the A14~\cite{ATL-PHYS-PUB-2014-021} set of tuned parameters and the \NNPDF[2.3lo] set of PDFs.


\section{Event reconstruction and object selection}
\label{sec:event_reco}

Events are required to have at least one $pp$ collision vertex candidate with at least two associated tracks with transverse momentum $\pT > \SI{0.5}{\GeV}$. The primary vertex is defined to be the vertex candidate with the largest scalar sum of transverse momenta of all associated tracks.
In this analysis, electrons, muons, and jets are the calibrated physics objects used.
For the charged leptons, two sets of quality and kinematic requirements are imposed, where the selection for signal leptons is tighter than for baseline leptons.

Electron candidates are reconstructed from energy deposits in the EM calorimeter matched to charged-particle tracks in the ID.
Baseline electrons are required to have $\pt > \SI{10}{\GeV}$ and to be reconstructed within $|\eta| < 2.47$, excluding the barrel--endcap transition region $1.37 < |\eta| < 1.52$. They must fulfil `loose' identification criteria, using a likelihood-based discriminant that combines information about tracks in the ID and energy deposits in the calorimeter system~\cite{EGAM-2018-01}, and are required to have a hit in the innermost layer of the pixel detector.
Furthermore, isolation requirements in both the calorimeter and the ID are imposed~\cite{EGAM-2018-01}.
An electron does not meet the isolation criteria if, after subtracting contributions from \pileup and the electron itself, the transverse energy deposited in the calorimeter within a surrounding cone of radius $\Delta R = 0.2$ exceeds \SI{20}{\percent} of the transverse energy of the electron.
Similarly, electron candidates are excluded if the scalar sum of the transverse momenta of tracks within a cone of radius $\Delta R = \min (\SI{10}{\GeV}/\pt(e), 0.2)$, excluding the track matched to the electron, is larger than \SI{15}{\percent} of the electron \pt.
In addition, each electron candidate's track must be matched to the primary vertex. This requires that the significance of its transverse impact parameter, $d_0$, satisfies $|d_0|/\sigma_{d_0} < 5$, where $\sigma_{d_0}$ is the uncertainty in $d_0$, and that the longitudinal distance $z_0$ from the primary vertex to the point where $d_0$ is measured satisfies $|z_0 \sin \theta| < \SI{0.5}{mm}$.
In order to suppress backgrounds due to hadrons misidentified as electrons, signal electrons must satisfy all baseline criteria and in addition `tight' identification criteria~\cite{EGAM-2018-01}, and have $\pT > \SI{28}{\GeV}$.

Muon candidates are reconstructed by combining
charged-particle tracks formed in the ID and in the muon spectrometer or by matching ID tracks to an energy deposit in the calorimeter compatible with a minimum ionising particle~\cite{MUON-2018-03}.
Baseline muons are required to have $\pT > \SI{10}{\GeV}$ and $|\eta| < 2.5$, and to satisfy the `loose' identification criteria~\cite{MUON-2018-03}.
Track-to-vertex matching is ensured by requiring the muon track to satisfy $|d_0|/\sigma_{d_0} < 3$ and $|z_0 \sin \theta| < \SI{0.5}{mm}$.
Signal muons must satisfy `medium' identification criteria and are required to have $\pT > \SI{28}{\GeV}$.
Additionally, signal muons must be isolated, requiring that the scalar \pt sum of all tracks within a cone of radius $\Delta R = \min\left(\SI{10}{\GeV}/\pt(\mu), 0.3\right)$ around the muon is less than \SI{6}{\percent} of the muon \pt.
 
Small-radius (small-$R$) jet candidates are built from particle-flow objects~\cite{PERF-2015-09,JETM-2018-05}, using the \antikt algorithm~\cite{Cacciari:2008gp,Fastjet} with a radius parameter of $R=0.4$.
The particle-flow algorithm combines information about tracks in the ID and energy deposits in the calorimeters to form the input for the jet reconstruction.
Jets are required to have $\pt > \SI{25}{\GeV}$ and $|\eta| < 2.5$.
To reject jets originating from \pileup interactions, jet candidates with $|\eta| < 2.4$ and $\pT < \SI{60}{\GeV}$ are required to satisfy the `tight' jet vertex tagger (JVT) criterion~\cite{ATLAS-CONF-2014-018}.
Small-$R$ jets containing a $b$-hadron decay are \btagged using a multivariate algorithm, called DL1r,
operating at a tagging efficiency of $77\,\%$ as determined in simulated \ttbar events~\cite{FTAG-2018-01,ATLAS:2022qxm}.

An overlap removal procedure is applied to prevent double counting of ambiguous reconstructed objects, using the baseline lepton definitions.
First, electron--muon overlap is handled by removing muons sharing a track in the ID with an electron if the muon is calorimeter-tagged, and otherwise removing the electron.
Subsequently, overlap between jets and leptons is removed by rejecting any jets within $\Delta R = 0.2$ of an electron and afterwards rejecting any electrons within $\Delta R = 0.4$ of a jet.
Similarly, jets are discarded if they have fewer than three associated tracks and are within $\Delta R = 0.2$ of a muon candidate.
Otherwise, the muon is rejected if it lies within $\Delta R = \text{min}(0.4, 0.04 + \SI{10}{\GeV}/\pt(\mu))$ of a jet.
 
The missing transverse momentum, with magnitude \met, is defined as the negative vectorial sum of the transverse momenta of all calibrated objects in an event, plus a track-based soft-term which takes into account energy depositions associated with the primary vertex but not with any calibrated object~\cite{PERF-2016-07}.

Finally, large-radius (large-$R$) jets are constructed from the selected small-$R$ jets using the \antikt algorithm with $R=1.0$. In order to reduce the impact of soft radiation, constituent small-$R$ jets with \pt less than \SI{5}{\percent} of the large-$R$ jet \pt are removed. These reclustered large-$R$ jets are required to have $\pt > \SI{150}{\GeV}$ and a mass larger than $\SI{50}{\GeV}$.


\section{Event selection and categorisation}
\label{sec:event_selection}

All events considered in this analysis must be selected by an \MET trigger. Since the trigger thresholds were raised during \RunTwo, a requirement of $\MET > \SI{250}{\GeV}$ is imposed to ensure full trigger efficiency over all data-taking periods. Events are also required to have exactly one signal lepton ($e$, $\mu$) and at least four small-$R$ jets of which at least one is $b$-tagged. A veto on a second lepton, fulfilling the baseline requirements, is used to suppress  \ttbar events with two leptons in the final state.
To reject events with \MET arising from mismeasured jets, the azimuthal angle between the missing transverse momentum vector \vecMET\ and
both the leading ($j_1$) and subleading ($j_2$) jets,
ordered in \pt, must satisfy the condition $|\Delta \phi (j_{i}, \vecMET)| > 0.4$, with $i \in \{1,2\}$.
In addition, events must have a transverse mass $\mtw > \SI{30}{\GeV}$, where $\mtw$ is defined as
\begin{align*}
\mtw  = \sqrt{2 \pt(\ell) \MET\left[1-\cos\left(\Delta \phi (\ell, \vecMET)\right)\right]}.
\end{align*}
 
After applying these requirements, referred to as `preselection' in Table~\ref{tab:control_validation_regions}, the dominant backgrounds come from \ttbar and \Wjets production.
A training region for the NNs is defined by applying further requirements listed in Table~\ref{tab:control_validation_regions}, which reduce the amount of background without decreasing the  sensitivity to the signal.
Requiring \mtw to be well above the $W$ boson mass peak, $\mtw > \SI{120}{\GeV}$, strongly reduces the \Wjets and semileptonic \ttbar background.
In these background processes the leptonic $W$ boson decay is the only source of \MET, while in the VLQ pair production additional
sources of \MET can arise from e.g., the $Z$ boson decay to neutrinos or the
$W$ boson decay to a hadronically-decaying tau-lepton and a neutrino.
The remaining \ttbar background originates from dileptonic \ttbar events where one lepton is not detected.
This type of \ttbar background is suppressed by requirements on the asymmetric transverse mass, \amtTwo~\cite{amt2,Lester:2014yga},
which is a variant of the \mttwo~\cite{mt2,Barr:2003rg} variable.
The latter is
applied to signatures where two or more particles are not detected directly (e.g.\ dileptonic \ttbar events, where the two neutrinos are not detected), and it is defined as
\begin{align*}
\mttwo = \min_{\vec{q}_{\mathrm{T}a}+\vec{q}_{\mathrm{T}b} = \vec{E}_{\mathrm{T}}^{\mathrm{miss}}}\left[ \max (m_{\mathrm{T}a},m_{\mathrm{T}b})\right].
\end{align*}
In this formula, $m_{\mathrm{T}a}$ and $m_{\mathrm{T}b}$ are transverse masses calculated using two sets of one or more visible particles, denoted by $a$ and $b$, respectively, and all possible combinations of missing transverse momenta $\vec{q}_{\mathrm{T}a}$ and $\vec{q}_{\mathrm{T}b}$, with $\vec{q}_{\mathrm{T}a}+\vec{q}_{\mathrm{T}b}=\vecMET$.
In the calculation of \amtTwo, the two sets of visible particles are asymmetric as they consist of the one identified signal lepton with one of the two jets
with highest $b$-tagging score on one hand and the other jet on the other.
Given the two possible lepton-jet pairings the combination with the lowest \amtTwo is taken.
For dileptonic \ttbar events, the \amtTwo distribution has a kinematic endpoint at the top-quark mass, while additional sources of \MET extend the distribution towards higher \amtTwo values. Events in the training region have to fulfil $\amtTwo>\SI{200}{\GeV}$.
At least one hadronic decay of a high-$\pT$ top quark or SM boson is expected for the considered signal.
Thus, at least one large-$R$ jet is required in the training region.
About 4\% of the simulated signal events are reconstructed in the training region
for $T\bar{T}$ or $B\bar{B}$ production with pure $T\rightarrow Zt$ and $B\rightarrow Wt$ decays and a VLQ mass of \SI{1.2}{\TeV}.

\begin{table}[tb]
\centering
\caption{
Overview of the event selections for the training region that is subdivided into a \lownnout control region and a signal region, for the top reweighting region, and for the control regions for \Wjets events and single-top-quark events.
}
\begin{tabular}{lcccc}
\toprule
\multicolumn{5}{c}{\textbf{Preselection}}                                                         \\ \midrule
\multicolumn{5}{c}{\MET triggers }                                                            \\
\multicolumn{5}{c}{$=1$ signal lepton}                                                          \\
\multicolumn{5}{c}{no additional baseline lepton}                                                     \\
\multicolumn{5}{c}{$\geq$ 4 jets}                                                             \\
\multicolumn{5}{c}{$\geq$ 1 $b$-jet}                                                           \\
\multicolumn{5}{c}{$\MET > \SI{250}{\GeV}$}                                                        \\
\multicolumn{5}{c}{$\mtw > \SI{30}{\GeV} $}                                                        \\
\multicolumn{5}{c}{$|\Delta \phi (j_{1,2}, \vecMET)| > 0.4$}                                               \\ \midrule
& \textbf{Training region} & \textbf{Top reweighting region} & $\boldsymbol{W}$\textbf{+jets CR} & \textbf{Single-top CR} \\
& \textbf{\lownnout CR\,/\,SR} &                                 &                                   &                        \\ \midrule
\mtw[\si{\GeV}]               &       $> 120$~~~~        &          $ > 120$~~~~           &          $\in [30,120]$           &     $\in [30,120]$     \\
\amtTwo[\si{\GeV}]            &       $> 200$~~~~        &           $<180$~~~~            &             $> 200$~~             &       $> 200$~~        \\
$b$-jet multiplicity           &        $\geq 1 $         &            $\geq 1 $            &              ~~$=1$               &       ~~$\geq 2$       \\
Large-$R$ jet multiplicity     &         $\geq 1$         &           $ \geq 1 $            &            ~~$\leq 1 $            &      ~~$\leq 1 $       \\
$m$(large-$R$ jet) [\si{\GeV}] &            --            &               --                &             $<150$~~              &        $<150$~~        \\
Lepton charge                  &            --            &               --                &             ~~~\,$+1$             &           --           \\
$\Delta R (b_{1},b_{2})$       &            --            &               --                &                --                 &        $> 1.4$~        \\
\nnout                         &    $< 0.5/ \geq 0.5$     &               --                &                --                 &           --           \\ \bottomrule
\end{tabular}
\label{tab:control_validation_regions}
\end{table}

\begin{figure}
\centering
\begin{subfigure}[b]{0.49\textwidth}
\centering
\includegraphics[width=0.9\textwidth]{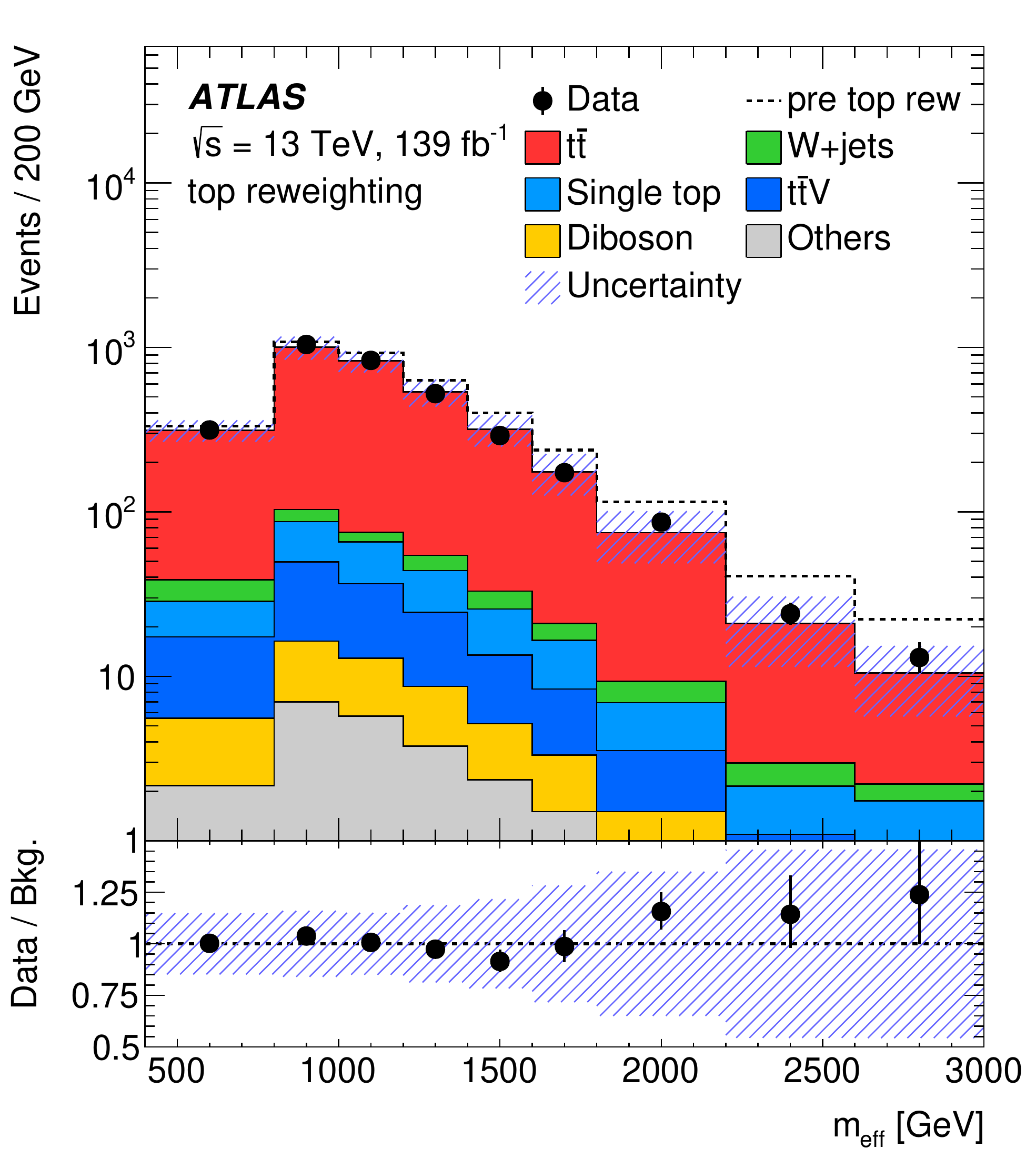}
\caption{}
\label{fig:selection:ttbar_rw_meff}
\end{subfigure}\begin{subfigure}[b]{0.49\textwidth}
\centering
\includegraphics[width=0.9\textwidth]{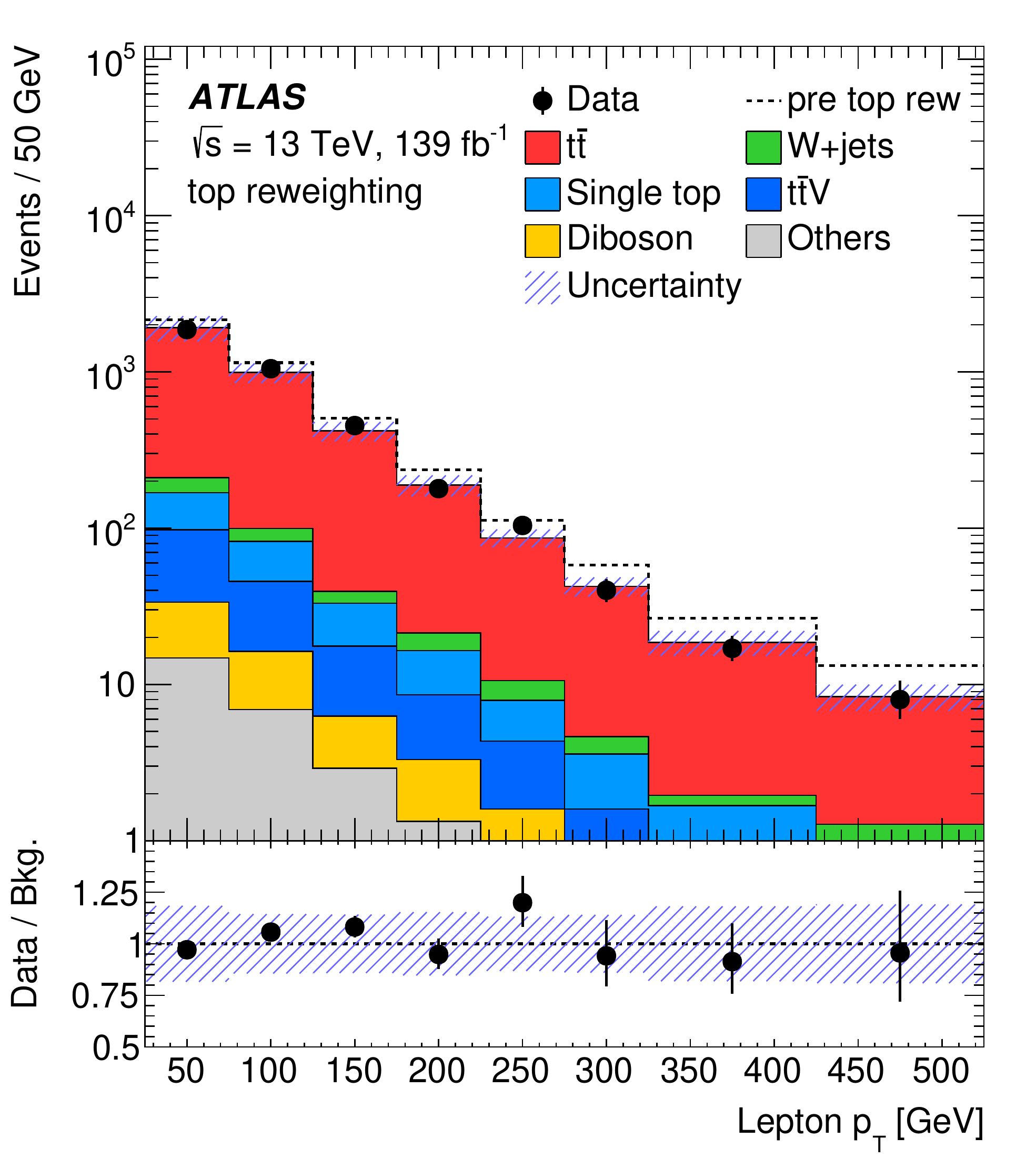}
\caption{}
\label{fig:selection:ttbar_rw_lepPt}
\end{subfigure}
\caption{Distributions of \subref{fig:selection:ttbar_rw_meff} \meff and \subref{fig:selection:ttbar_rw_lepPt} lepton \pT in the top reweighting region after applying the reweighting factors to the \ttbar and single-top background. The dashed line indicates the total background before the reweighting. The band includes statistical and systematic uncertainties. Minor background contributions from \ttH, \tWZ, and \Zjets are combined into \textit{Others}. The ratios of the data to the expected background are shown in the bottom panels of the plots. The last bin in each distribution contains the overflow.}
\label{fig:selection:ttbar_rw}
\end{figure}

The \ttbar background, which is a major background in this analysis, is not modelled accurately at high transverse momenta~\cite{TOPQ-2018-17,TOPQ-2018-15}.
Therefore, a reweighting procedure, referred to as `top reweighting' in the following, is applied. Reweighting factors are derived in bins of the jet multiplicity (4, 5, 6, ${\geq}7$) as a function of the effective mass \meff,
defined as the scalar sum of the \pt of all reconstructed objects and \MET. The reweighting factors are determined for the sum of the \ttbar and single-top
backgrounds
using their nominal prediction
and are parameterised with a linear function.   
They are derived
from a comparison between data and MC simulation
in a dedicated top reweighting region (see Table~\ref{tab:control_validation_regions}), which is defined in the same way as the training region except for an inverted 
\amtTwo requirement.
This requirement is strengthened to $\amtTwo < \SI{180}{\GeV}$ in order to have a higher \ttbar purity of about 90\% and less signal contamination in the tails of the \meff distribution.
The resulting reweighting factors are applied to \ttbar and single-top-quark events in each of the defined analysis regions, which changes the event yields and leads to an improved modelling. This can be seen in Figure~\ref{fig:selection:ttbar_rw} which compares the data with MC simulation in the top reweighting region after reweighting, and also shows the MC expectation before reweighting.

\begin{figure}
\centering
\begin{subfigure}[b]{0.49\textwidth}
\centering
\includegraphics[width=0.9\textwidth]{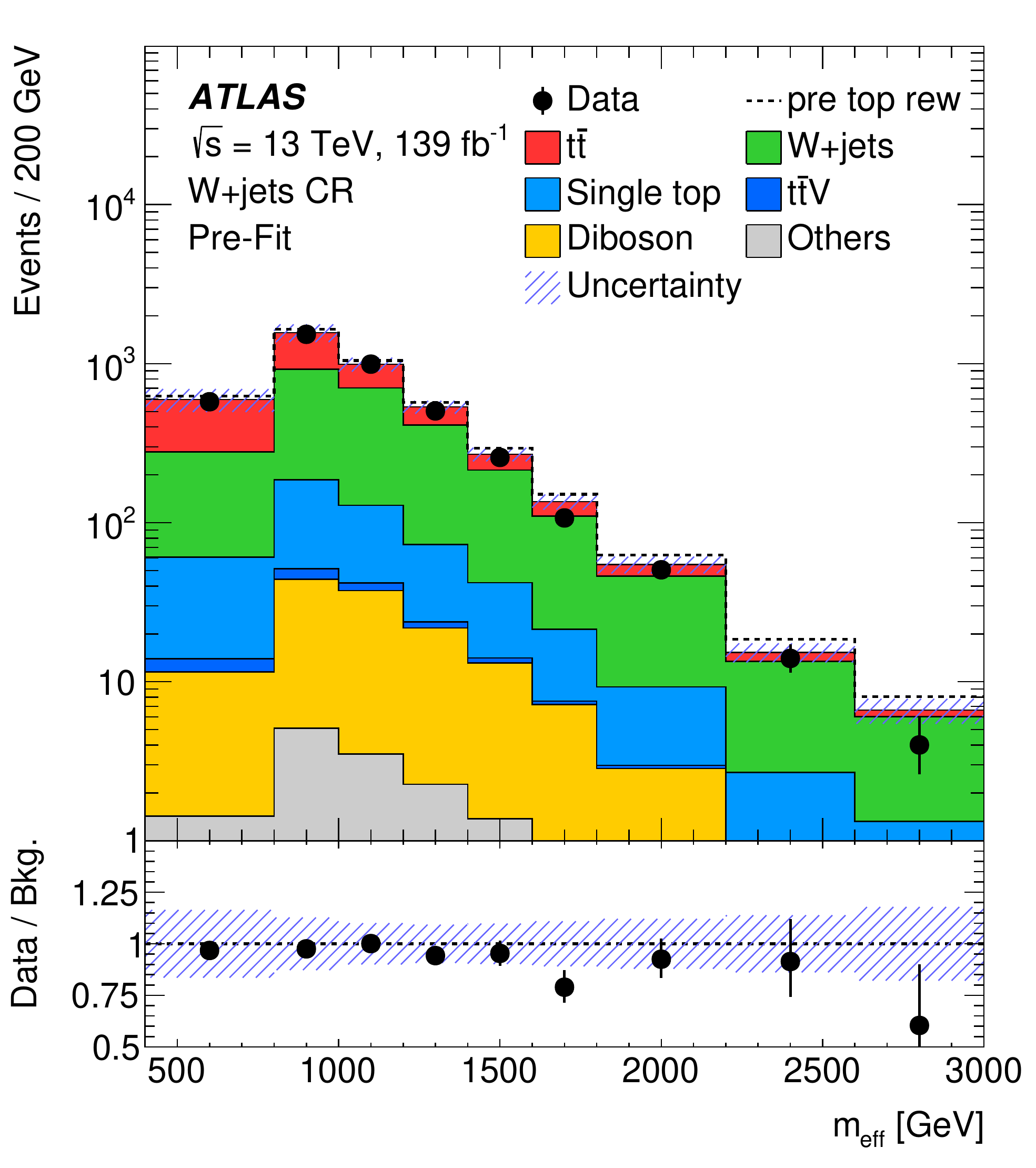}
\caption{}
\label{fig:selection:wjetsCR_meff}
\end{subfigure}\begin{subfigure}[b]{0.49\textwidth}
\centering
\includegraphics[width=0.9\textwidth]{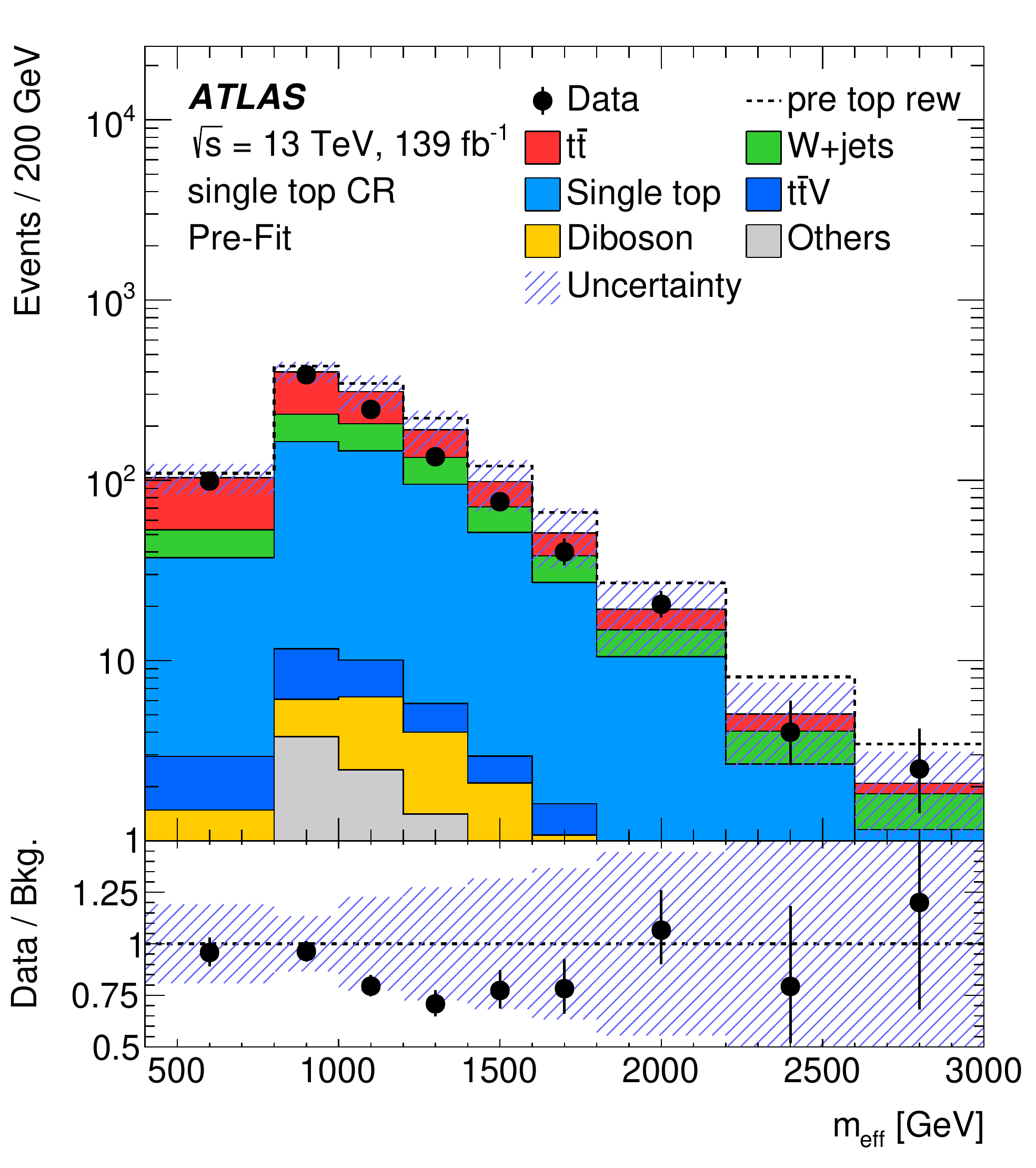}
\caption{}
\label{fig:selection:stCR_lepPt}
\end{subfigure}
\caption{Distributions of \meff in \subref{fig:selection:wjetsCR_meff} the \Wjets CR and \subref{fig:selection:stCR_lepPt} the single-top CR after applying the top reweighting factors to the simulated \ttbar and single-top-quark events.  The dashed line indicates the total background before the reweighting. Minor background contributions from \ttH, \tWZ, and \Zjets are combined into \textit{Others}. The band includes statistical and systematic uncertainties. The ratios of the data to the expected background are shown in the bottom panels of the plots. The last bin in each distribution contains the overflow.}
\label{fig:meff_wjetsCR_stCR}
\end{figure}

Control regions are defined for the \Wjets and single-top-quark backgrounds so as to be enriched in the respective background and have negligible contamination from signal.
Both CRs are defined to be orthogonal to the top reweighting region and the training region by modifying the requirement on \mtw, using a window of $\SI{30}{\GeV}<\mtw < \SI{120}{\GeV}$ around the $W$ boson mass. In order to reduce the \ttbar background contribution in these regions, the large-$R$ jet multiplicity is required to be less than two, and if a large-$R$ jet is present its mass has to be below \SI{150}{\GeV}.
For the \Wjets CR, the contribution from \ttbar events is further reduced by selecting only events with exactly one $b$-tagged jet. Since the cross-section for $W^+$ production is larger than for $W^-$ production, higher \Wjets purity is achieved by selecting only events with a positively charged lepton. In the single-top CR, the contribution from \Wjets is reduced by requiring at least two $b$-tagged jets with an angular separation of $\Delta R (b_{1},b_{2})> 1.4$ between the two highest-\pt $b$-jets.
Table~\ref{tab:control_validation_regions} summarises the selection criteria for both CRs, and Figure~\ref{fig:meff_wjetsCR_stCR} compares the effective mass distribution in data with that in MC simulation after top reweighting, and also shows the MC expectation before reweighting.
 
\section{Neural network training}
\label{sec:nn_training}

\label{nns}
 
To enhance the separation between signal and background events, NNs with several input variables combined into a single discriminant are employed. They are trained for various signal hypotheses using the simulated signal and background events in the training region. For $T\bar{T}$ production, four NNs are trained for different branching ratios $\mathcal{B}(Zt,Ht,Wb)$ covering the
region of the branching ratio plane where this analysis is sensitive: $(0.8,0.1,0.1)$, $(0.2,0.4,0.4)$, $(0.4,0.1,0.5)$, $(0.4,0.5,0.1)$. Similarly, three NNs are trained for $B\bar{B}$ production, considering the branching ratios $\mathcal{B}(Zb,Hb,Wt) =$ $(0.1,0.1,0.8)$, $(0.4,0.1,0.5)$, and $(0.1,0.4,0.5)$.
 
\begin{figure}
\centering
\begin{subfigure}[b]{0.49\textwidth}
\centering
\includegraphics[width=0.9\textwidth]{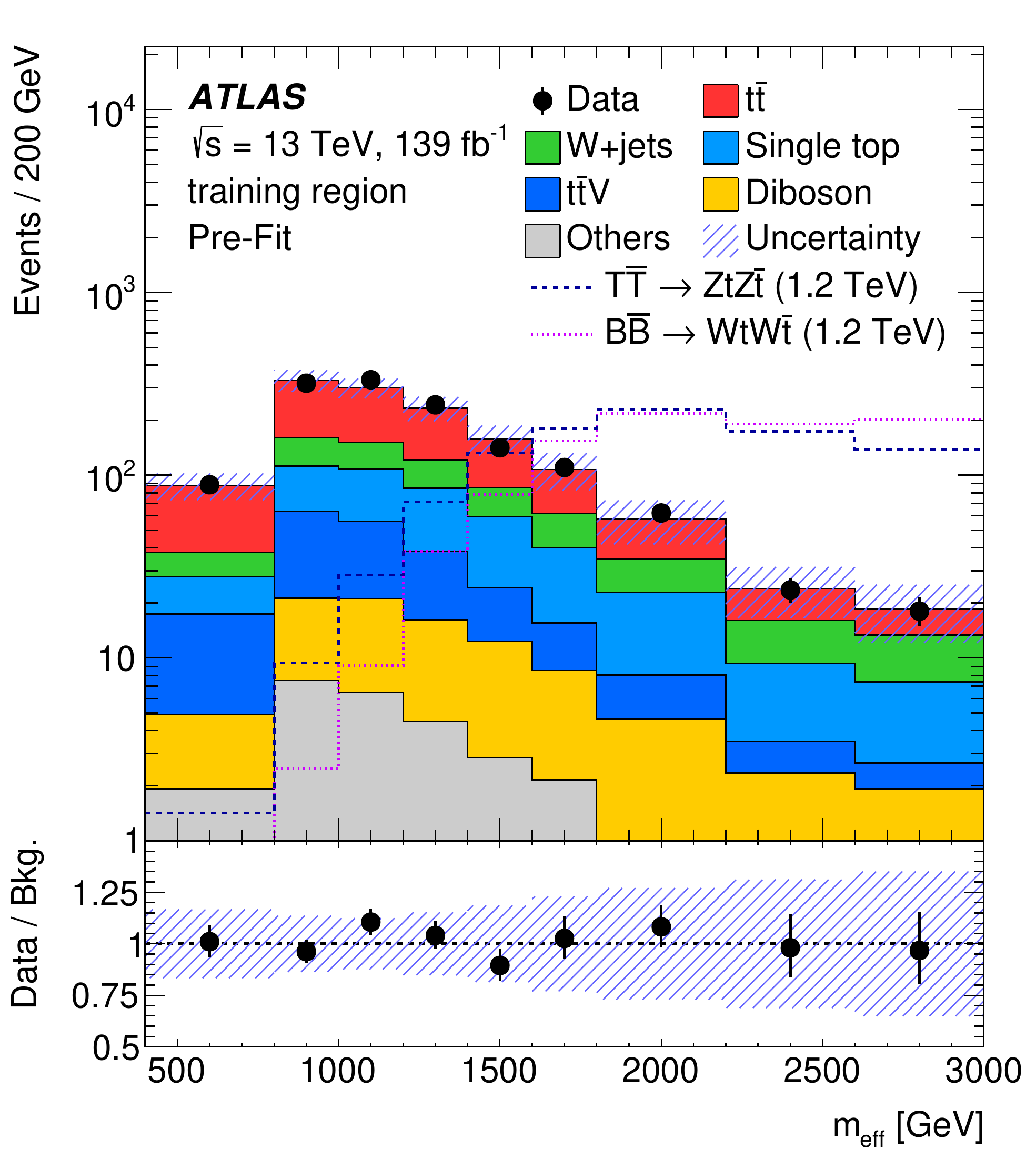}
\caption{}
\label{fig:selection:TR_meff}
\end{subfigure}\begin{subfigure}[b]{0.49\textwidth}
\centering
\includegraphics[width=0.9\textwidth]{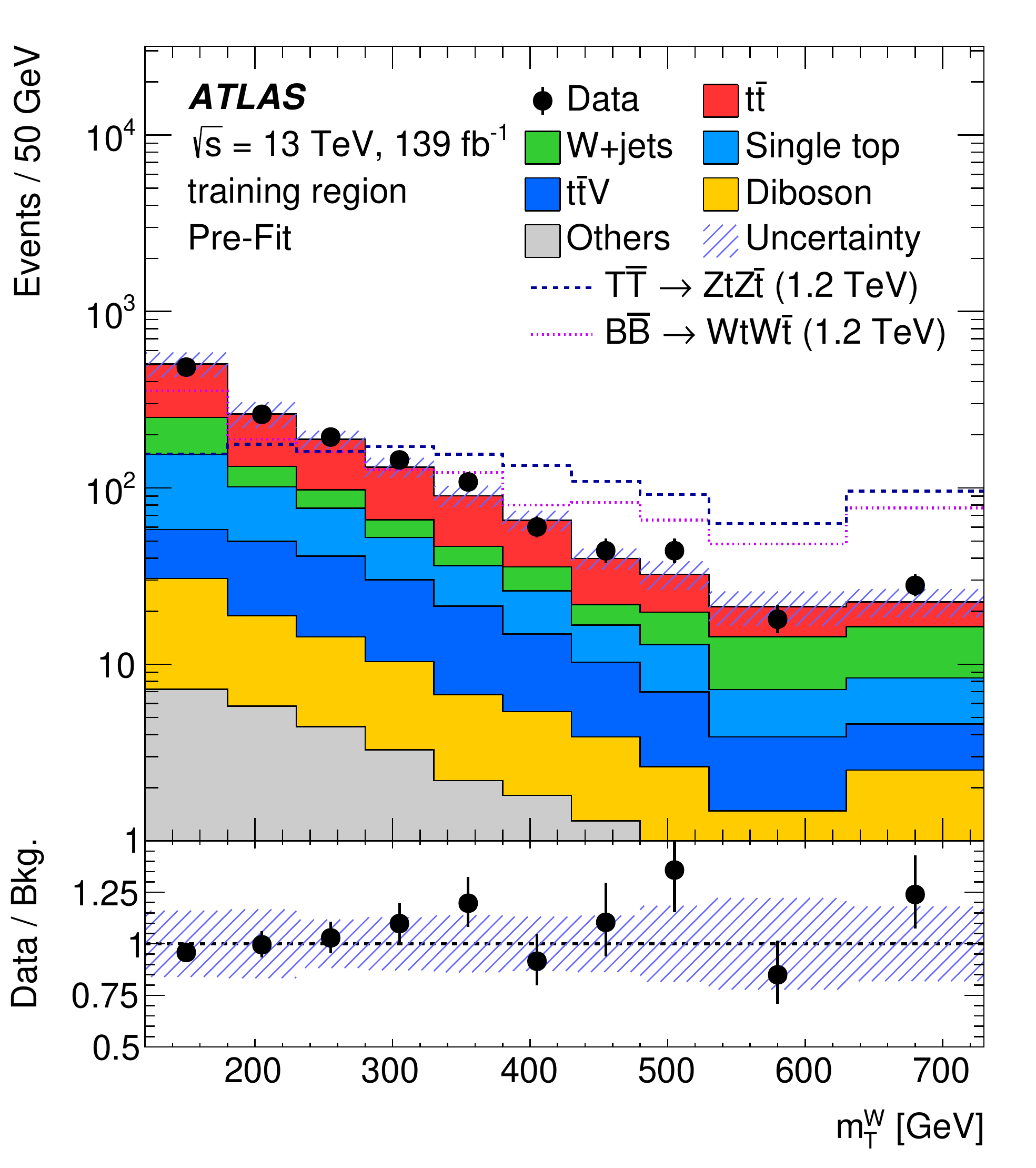}
\caption{}
\label{fig:selection:TR_mtw}
\end{subfigure}
 
\begin{subfigure}[b]{0.49\textwidth}
\centering
\includegraphics[width=0.9\textwidth]{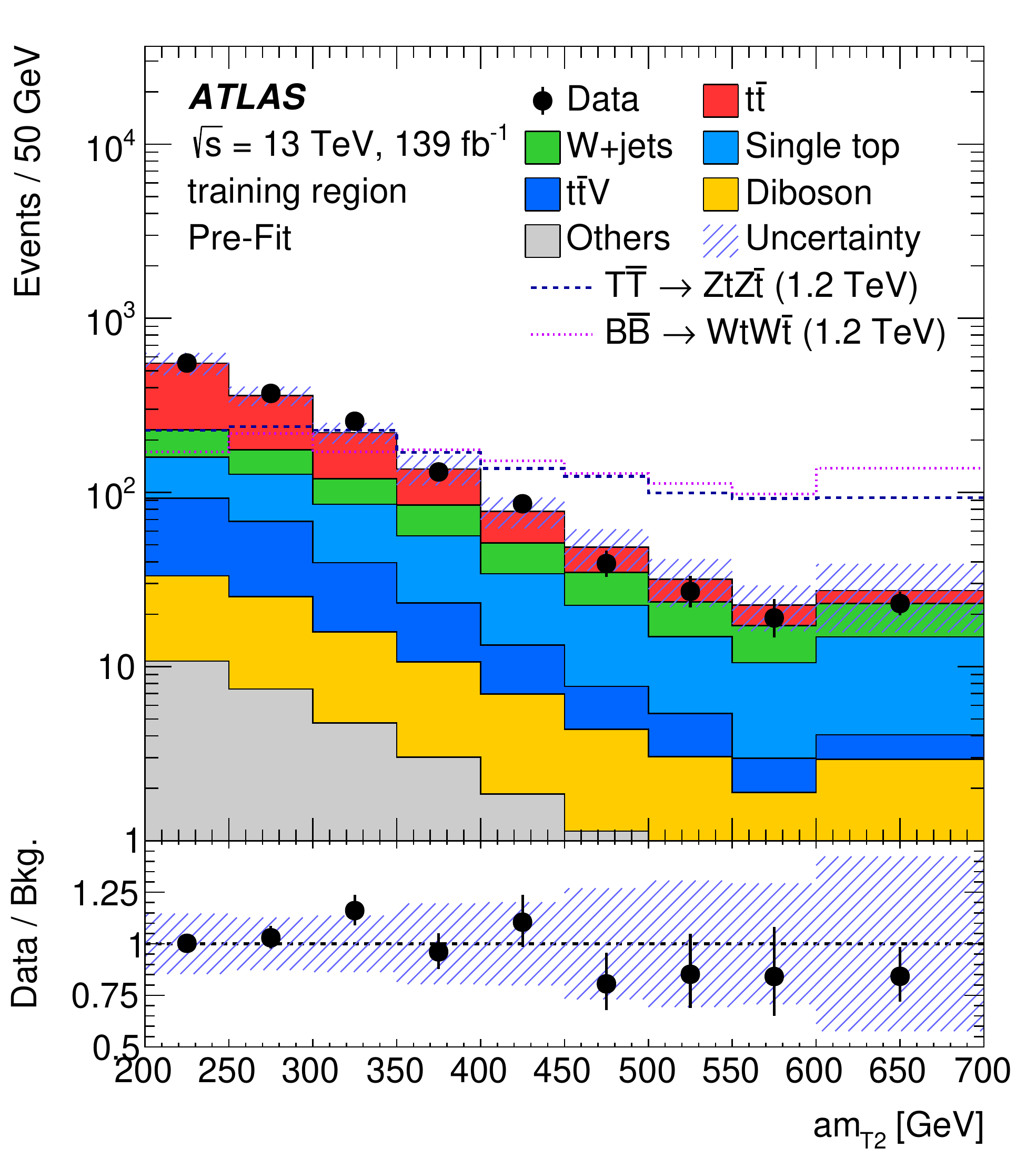}
\caption{}
\label{fig:selection:TR_amt2}
\end{subfigure}\begin{subfigure}[b]{0.49\textwidth}
\centering
\includegraphics[width=0.9\textwidth]{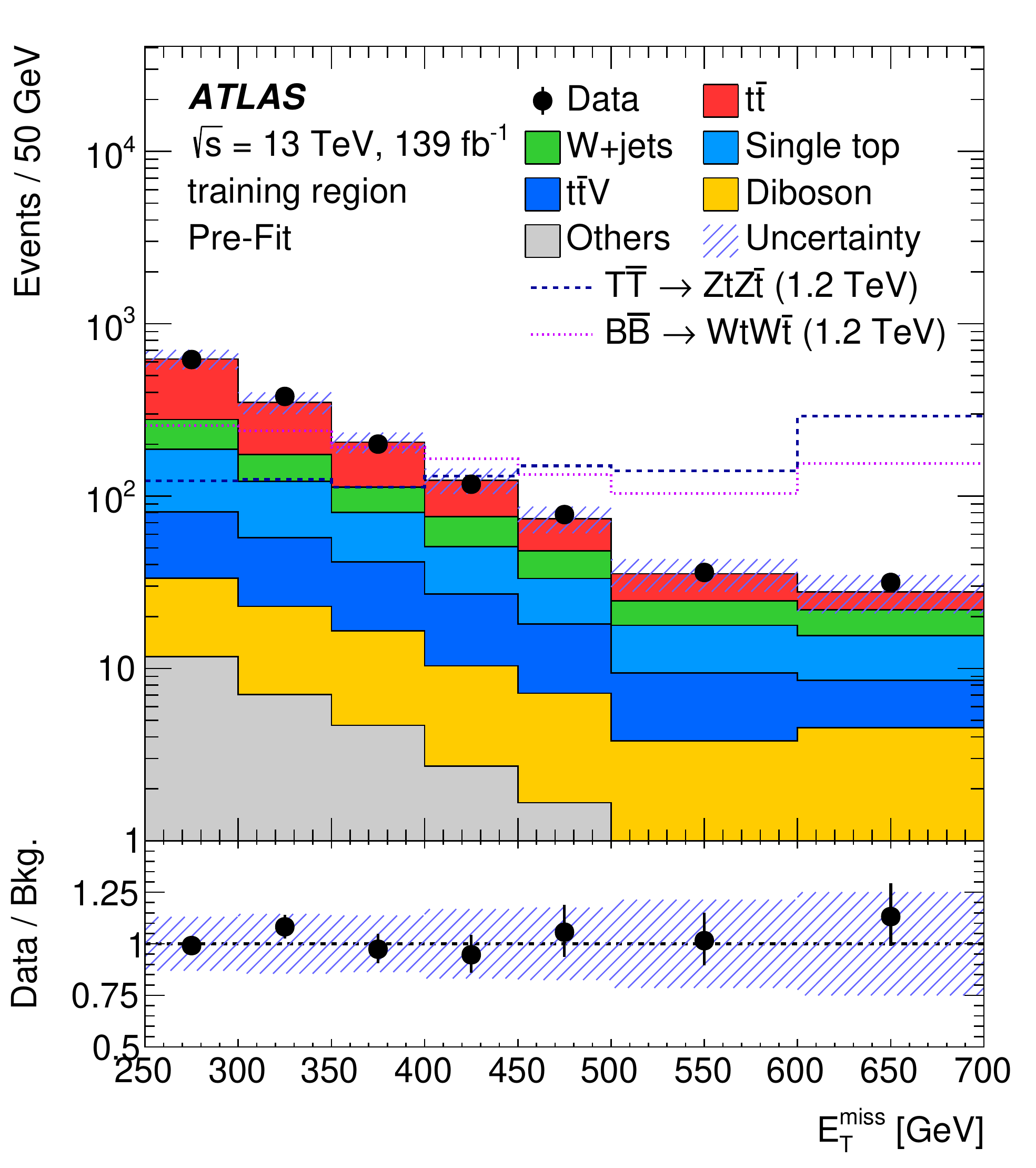}
\caption{}
\label{fig:selection:TR_met}
\end{subfigure}
\caption{Distributions of NN input variables in the training region: \subref{fig:selection:TR_meff} \meff,  \subref{fig:selection:TR_mtw} \mtw, \subref{fig:selection:TR_amt2} \amtTwo and \subref{fig:selection:TR_met} \met. Minor background contributions from \ttH, \tWZ, and \Zjets are combined into \textit{Others}. The signal distributions for $T\bar{T}\rightarrow ZtZ\bar{t}$ and $B\bar{B}\rightarrow WtW\bar{t}$, assuming a VLQ mass of \SI{1.2}{\TeV}, are overlaid and normalised to the total background prediction. The band includes statistical and systematic uncertainties. The ratios of data to the expected background are shown in the bottom panels of the plots. The last bin in each distribution contains the overflow.}
\label{fig:input-vars}
\end{figure}

The NNs are implemented using the NeuroBayes package~\cite{feindt-2004,Feindt:2006pm}, which combines a
three-layer feed-forward NN with
preprocessing of the input variables prior to their presentation to the NN.
The main purpose of the preprocessing is to facilitate optimal network training by ordering the input variables according to their separation power, taking correlations into account, and removing all but the most relevant ones.
Sets of input variables are selected for their ability to discriminate between signal and background.
Table~\ref{tab:analysis:inputvariables} lists the input variables that are used to train at least one NN.
Each set of input variables is composed of observables reflecting the signal topology, e.g.\ the high VLQ mass via \meff or the properties of the reclustered large-$R$ jets.
Other important variables are the $b$-jet multiplicity and the transverse masses, \mtw and \amtTwo, that are used to define the CRs and training region.
It is checked that all input variables are modelled well.
As an example, the distributions of four important input variables in the training region are shown in \Fig{\ref{fig:input-vars}}.

\begin{table}
\centering
\caption{Input variables to the NN training, sorted in descending discriminating power
between signal and background. The order is not strict as it depends on the
VLQ type and branching ratio the NN is trained for.}
\label{tab:analysis:inputvariables}
\begin{tabular}{ll}
\toprule
Variable            & Description                                                           \\ \midrule
$\meff$             & scalar sum of the transverse momenta of leptons, jets, and \MET       \\
$N_{b\text{-jets}}$ & $b$-jet multiplicity                                                  \\
$\mtw$              & transverse mass of lepton and \met                                    \\
$\amtTwo$           & asymmetric transverse mass                                            \\
\RCjetpttwo         & transverse momentum of second-highest-\pt large-$R$ jet                   \\
\deltaphijetoneMET  & azimuthal angle between \met and highest-\pt jet           \\
\MET                & missing transverse momentum                                           \\
\jetetaone          & pseudorapidity of highest-\pt jet                                      \\
\RCjetmassone       & mass of highest-\pt large-$R$ jet                                     \\
\nconstRCone        & number of small-$R$ jets reclustered to the highest-\pt large-$R$ jet \\
\leppt              & transverse momentum of lepton                                         \\
\jetptthree         & transverse momentum of third-highest-\pt jet                          \\
\jetpttwo           & transverse momentum of second-highest-\pt jet                             \\ \bottomrule
\end{tabular}
\end{table}

NeuroBayes uses Bayesian regularisation techniques for the training process to improve the generalisation performance and to avoid overtraining.
In general, the network infrastructure consists of one input node for each input variable, plus one bias node, an arbitrary, user-defined number of hidden nodes, and one output node which gives a continuous NN output score (\nnout) in the interval $(0,+1)$, where \nnout values close to zero indicate background-like events and values close to one correspond to signal-like events.
For the NNs in this analysis, 15 nodes are used in the hidden layer and the ratio of signal to background events in the training is chosen to be 1:1.
The different background processes are weighted according to their expected event contribution.
All the main backgrounds, \ttbar, \Wjets, single top quark, and \ttV, are used in the training. For the signal process, VLQ masses from \SI{1}{\TeV} to \SI{1.5}{\TeV} are combined in each training.
Events at different signal masses enter with the same cross-section
when composing the training sample in order to prevent the lower masses with higher cross-sections from dominating.
As a check for potential overtraining,
only \SI{80}{\percent} of the simulated events serve as input to the training, while the remaining
\SI{20}{\percent} are used as a test sample.
No signs of overtraining are observed.
After the training step, all simulated signal and background events, as well as the observed data events, are processed by the NNs in order to
get an \nnout value for each event.
For each NN, the training region is divided into a \lownnout CR with $\nnout < 0.5$, and the SR with $\nnout > 0.5$.


\section{Systematic uncertainties}
\label{sec:syst_uncertainties}

Several sources of experimental and theoretical systematic uncertainty are considered.
The experimental uncertainties are mainly related to the reconstruction and calibration of the final-state physics objects,
while the theoretical uncertainties are associated with the modelling of the various processes by the MC event generators.
The sources of the largest systematic uncertainties in the analysis are related to the modelling
of the major background processes and 
to the jet energy resolution.

For \ttbar and single-top production the following systematic uncertainties
related to the event modelling
are applied. The uncertainty in the matching procedure between the ME generator and parton shower is assessed by comparing the nominal \POWPY[8] samples with alternative samples generated by \MGNLO and showered by \PYTHIA[8].
In order to estimate the uncertainties in the modelling of
the underlying event, the parton shower, and the hadronisation, the
nominal samples are compared with a \POWHER[7]~\cite{Bellm:2017jjp} prediction.
Uncertainties related to the choice of renormalisation and factorisation scales of the matrix-element calculation are considered by independently doubling and halving the scales.
The impact of initial-state radiation (ISR) is estimated by varying $\alphas$ in the A14 tune.
Similarly, the uncertainty related to final-state radiation (FSR) is assessed by varying the renormalisation scale for final-state parton-shower emissions by a factor of two.
The uncertainty related to the choice of scale for the matching of the matrix-element calculation for the \ttbar process to the parton shower is evaluated by comparing the nominal samples with an alternative sample produced with the \hdamp parameter set to $\hdamp = 3.0 \, m_t$.
Uncertainties due to PDFs are obtained using the PDF4LHC15 combined PDF set~\cite{Butterworth:2015oua}. A dominant systematic uncertainty in the modelling of the single-top processes stems from the handling of the interference between \ttbar and \tW at NLO. The uncertainty is estimated by comparing the nominal sample for \tW production generated using the diagram-removal scheme with an alternative sample using the diagram-subtraction scheme~\cite{Frixione:2008yi,Re:2010bp}.
Finally, an additional \SI{30}{\percent} normalisation uncertainty is assigned to events from \ttbar + heavy-flavour jets production~\cite{HIGG-2017-03}.

Uncertainties in the \topreweighting procedure arise from the chosen form of the parameterised function and the statistical uncertainties of events in the reweighting region. These are accounted for by varying the parameterised function by $\pm 1\sigma$ from its nominal value, using the uncertainties of the fit parameters and taking their correlations into account. Each of the four jet bins for which the reweighting is determined is treated as an independent source of uncertainty.
 
For all other considered processes, namely \Vjets, diboson, \ttV, \ttH, and \tWZ production, the renormalisation and factorisation scales are independently varied by a factor of two.
A \SI{30}{\percent} uncertainty is assigned to the heavy-flavour component of the \Wjets process, based
on a comparison between \SHERPA[2.2.1] and data~\cite{HIGG-2018-51}.

Backgrounds without a free-floating normalisation parameter in the profile-likelihood fit are assigned a theoretical cross-section uncertainty. For the \ttH process, an \SI{11}{\percent}~\cite{deFlorian:2016spz} uncertainty is assigned and for \ttV and \tWZ the uncertainty amounts to \SI{15}{\percent} and \SI{12}{\percent}~\cite{deFlorian:2016spz}, respectively. The cross-section uncertainty is taken to be \SI{6}{\percent}~\cite{ATL-PHYS-PUB-2016-002}  for diboson production and \SI{5}{\percent}~\cite{STDM-2015-03} for the \Zjets process.
 
Besides theoretical systematic uncertainties, detector-related uncertainties are considered in the analysis, the dominant one being the
jet energy resolution~\cite{JETM-2018-05}.
Additional jet-related uncertainties are due to the jet energy scale, the jet mass scale and resolution, the efficiency of the JVT requirements, and the $b$-jet identification~\cite{FTAG-2018-01,FTAG-2020-08}. Uncertainties associated with leptons arise from the efficiencies of the lepton identification, isolation, and reconstruction, as well as the lepton energy scale and resolution~\cite{EGAM-2018-01,MUON-2018-03}. Further experimental uncertainties are related to the scale and resolution of the track soft-term in the \met calculation~\cite{ATLAS-CONF-2018-023}. Additional contributions to the total systematic uncertainty come from the uncertainty in the integrated luminosity and the \pileup profile.


\section{Statistical analysis}
\label{sec:stat_interpretation}

The 
signal-enriched part of the binned NN output distribution (\nnout > 0.5), and the overall number of events in the \lownnout, \Wjets, and single-top CRs are used to test for the presence of a signal.
For hypothesis testing, binned profile-likelihood fits are performed for each of the seven NNs separately, following a modified frequentist method~\cite{Read:2002hq} implemented in \textsc{RooStats}~\cite{RooFit}, and taking the systematic uncertainties affecting the signal and background expectations into account as nuisance parameters.

The binned likelihood function $\mathcal{L}(\mu,\theta)$ is constructed as the product of Poisson probability terms
over all bins. 
It depends on the signal strength parameter $\mu$, a factor multiplying the theoretical signal production cross-section, and $\theta$, a set of nuisance parameters, constrained in the likelihood function by Gaussian or log-normal priors.
The \lownnout, \Wjets, and single-top CRs are used to mainly control the normalisations of \ttbar, \Wjets, and single-top backgrounds, for which additional unconstrained normalisation factors ($\mu_{t\bar{t}}$, $\mu_\text{$W$\!+jets}$, and $\mu_\text{single top}$) are included in the likelihood function. The number of events expected in a bin depends on the normalisation factors as well as on the nuisance parameters.
The nuisance parameters $\theta$ adjust the expectations for signal and background according to the corresponding systematic uncertainties, and their fitted values correspond to the amounts that best fit the data.

In order to avoid double-counting of normalisation uncertainties for the free-floating background processes, only shape effects and acceptance differences between the CRs and the SR are included when considering the systematic uncertainties in their modelling.
A smoothing algorithm is applied to the templates for the systematic variations in case the statistical fluctuations between bins in the signal region are large.
Furthermore, the templates for all systematic variations are symmetrised.
Some of the dominant systematic uncertainties related to the modelling of the major background processes are one-sided
and are symmetrised by mirroring the uncertainty.
To simplify the fitting procedure, nuisance parameters are only included for systematic uncertainties that affect the event yield by more than \SI{1}{\percent}  for a process in any bin.
Normalisation and shape components for a source of systematic uncertainty are treated separately in this
procedure.

The test statistic $q_{\mu}$ is defined as the profile likelihood ratio, $q_{\mu} = -2 \ln{\mathcal{L}(\mu,\hat{\hat{\theta}})/\mathcal{L}(\hat{\mu},\hat{\theta})}$,
where $\hat{\mu}$ and $\hat{\theta}$ are the values of the parameters that simultaneously maximise the likelihood function,
and $\hat{\hat{\theta}}$ are the values of the nuisance parameters that maximise the likelihood function for a fixed value of $\mu$.
The compatibility of the observed data with the background-only hypothesis is tested by setting $\mu = 0$ in the test statistic $q_0$.
Upper limits on the signal production cross-section for each considered signal scenario are computed using $q_\mu$ in the CL$_\text{s}$ method~\cite{Read:2002hq} with the asymptotic approximation~\cite{Cowan:2010js}.
A given signal scenario is considered to be excluded at $\geq 95\%$ confidence level (CL) if the value of the
signal production cross-section (parameterised by $\mu$) yields a CL$_\text{s}$ value $\leq 0.05$.


\section{Results}
\label{sec:results}

\begin{figure}
\centering
\begin{subfigure}[b]{0.33\textwidth}
\centering
\includegraphics[width=\textwidth]{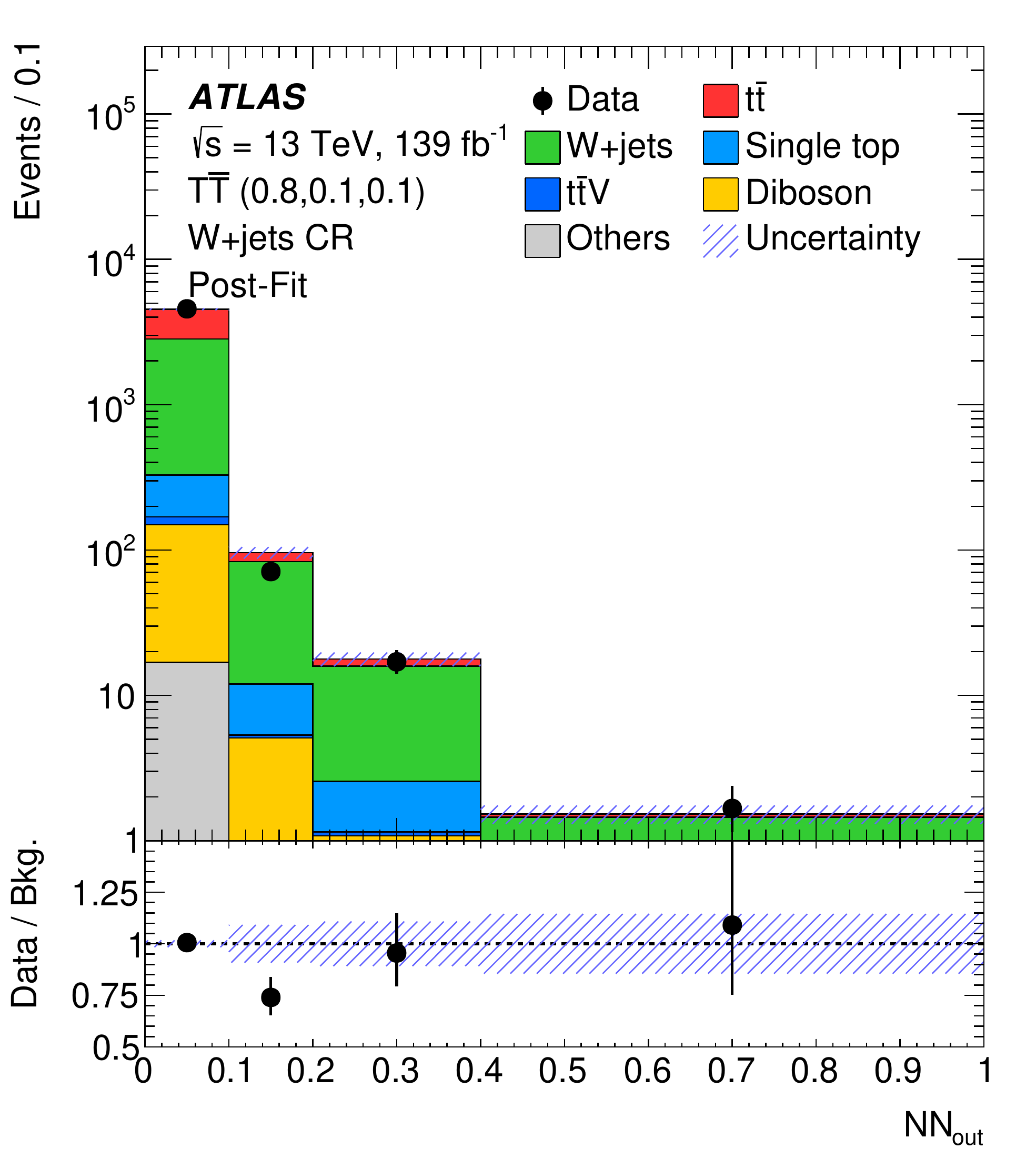}
\caption{}
\label{fig:results:NN_control:Wjets}
\end{subfigure}\begin{subfigure}[b]{0.33\textwidth}
\centering
\includegraphics[width=\textwidth]{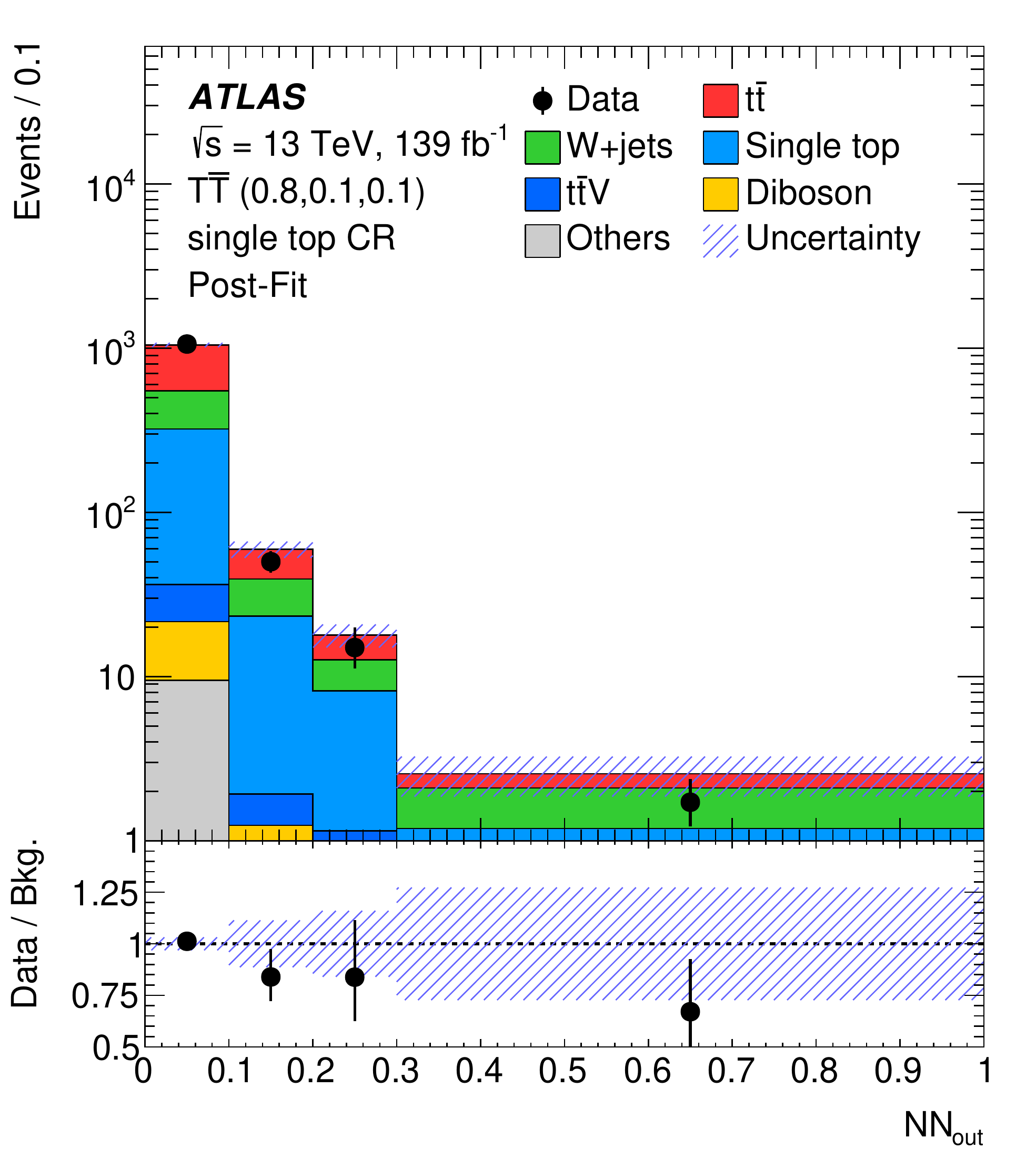}
\caption{}
\label{fig:results:NN_control:singletop}
\end{subfigure}\begin{subfigure}[b]{0.33\textwidth}
\centering
\includegraphics[width=\textwidth]{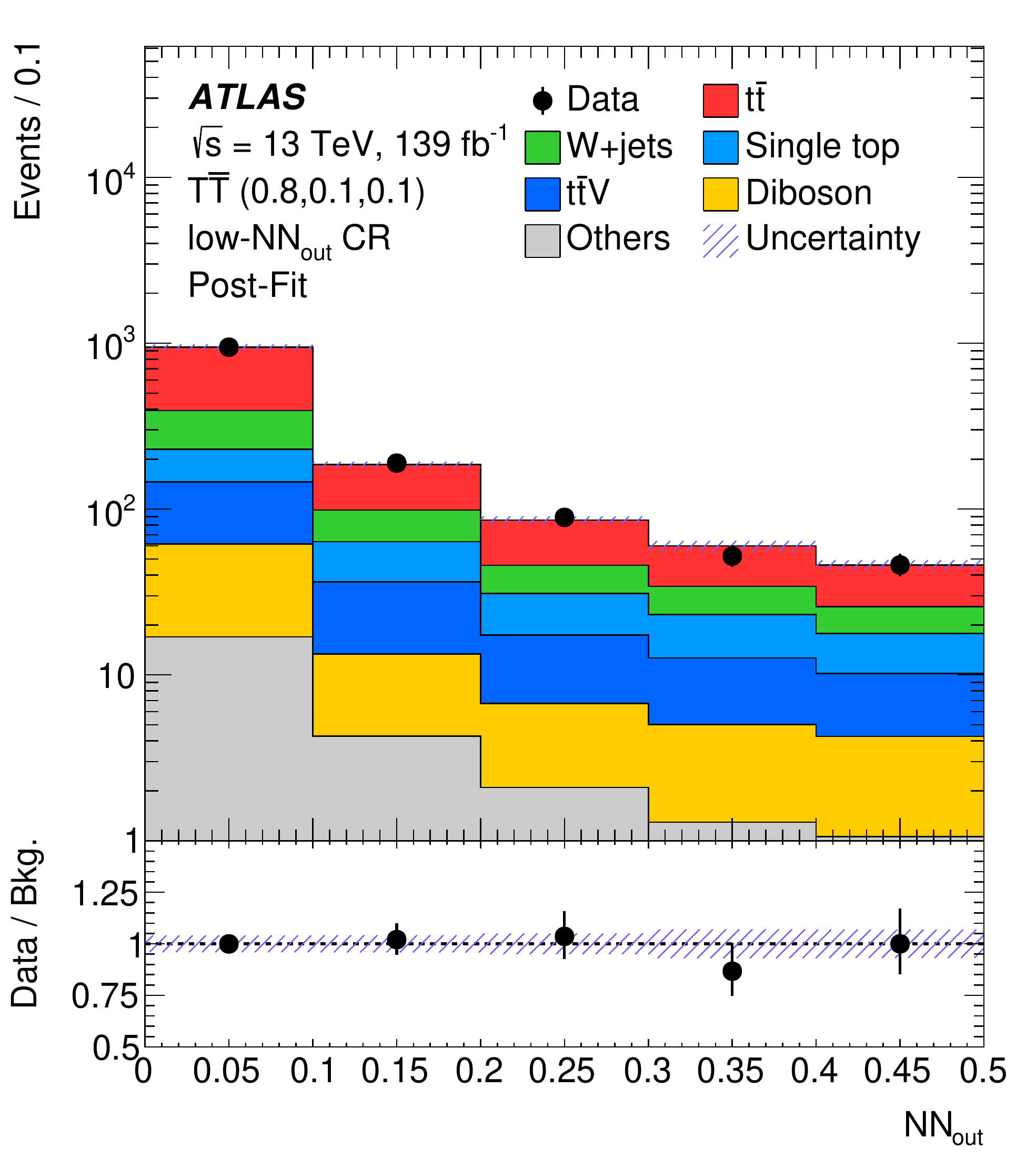}
\caption{}
\label{fig:results:NN_control:lowNNout}
\end{subfigure}
 
\begin{subfigure}[b]{0.33\textwidth}
\centering
\includegraphics[width=\textwidth]{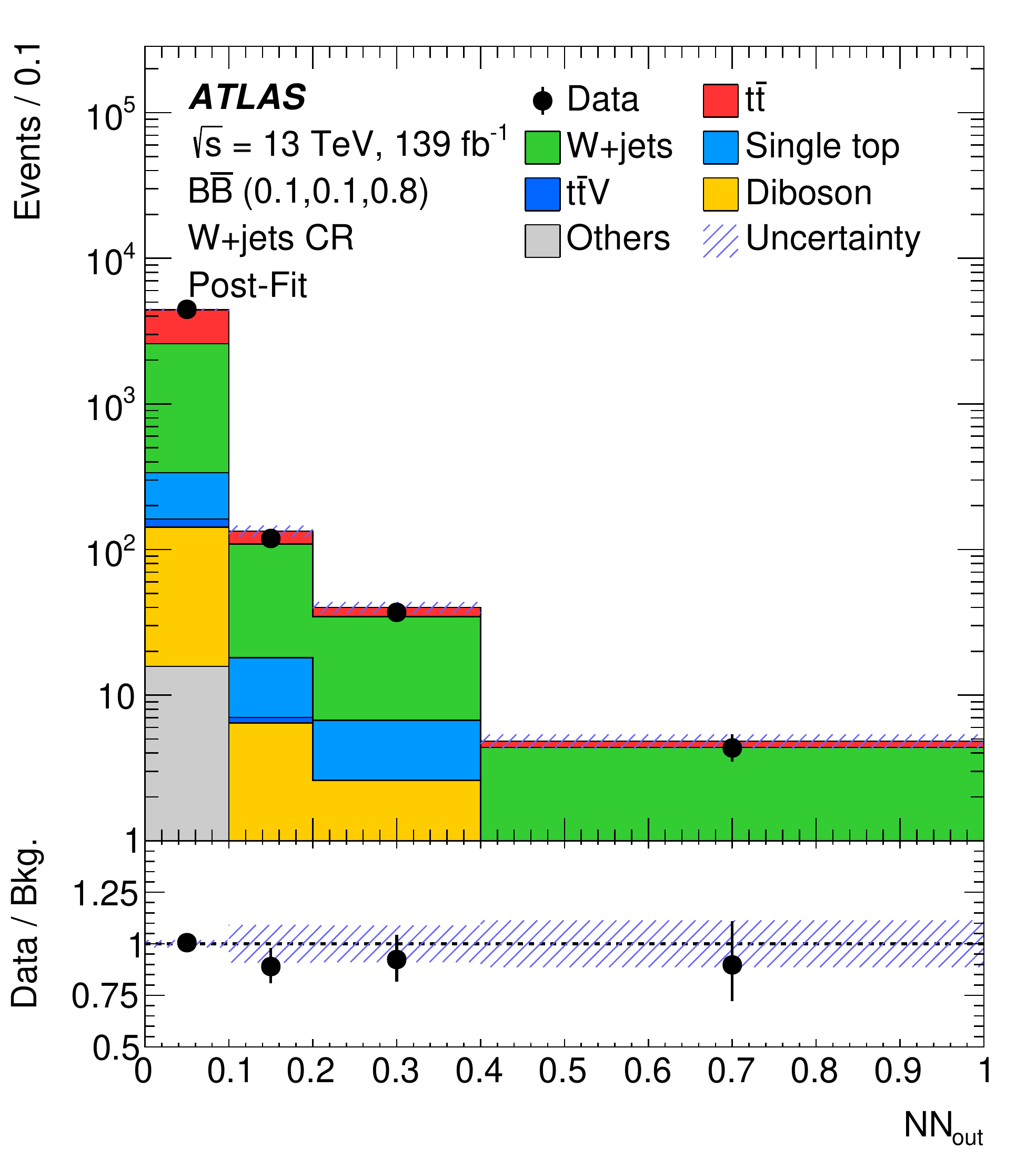}
\caption{}
\label{fig:results:NN_control4:Wjets}
\end{subfigure}\begin{subfigure}[b]{0.33\textwidth}
\centering
\includegraphics[width=\textwidth]{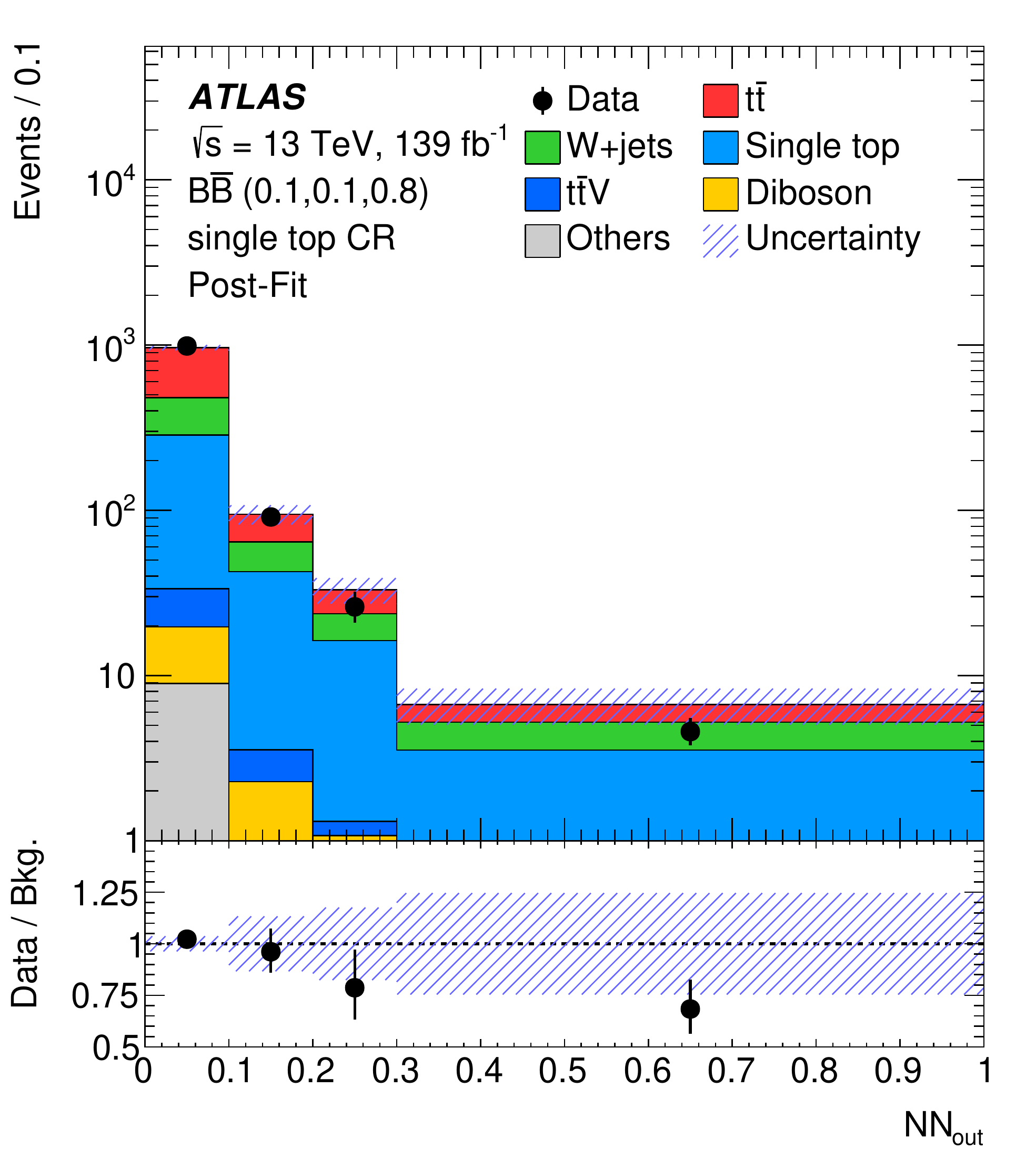}
\caption{}
\label{fig:results:NN_control4:singletop}
\end{subfigure}\begin{subfigure}[b]{0.33\textwidth}
\centering
\includegraphics[width=\textwidth]{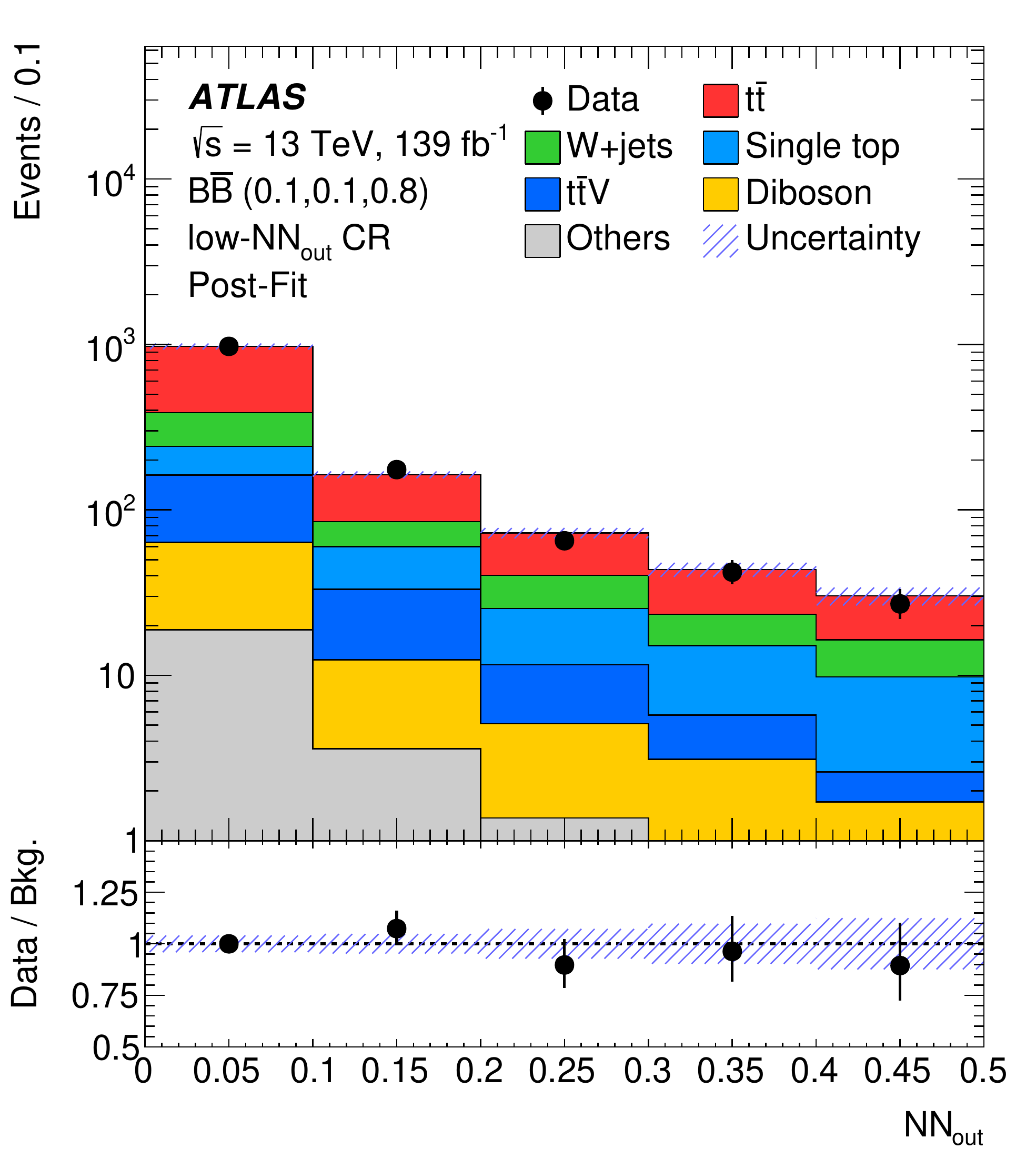}
\caption{}
\label{fig:results:NN_control4:lowNNout}
\end{subfigure}
\caption{Data and background expectation in the 
\Wjets CR (left panels),
the single-top CR (middle panels), and
the \lownnout CR (right panels) after a background-only fit to data (Post-Fit) for a NN training considering a VLT signal with $\mathcal{B}(Zt,Ht,Wb) = (0.8,0.1,0.1)$ (upper panels)
and a VLB signal with $\mathcal{B}(Zb,Hb,Wt) = (0.1,0.1,0.8)$ (lower panels). Minor background contributions from \ttH, \tWZ, and \Zjets are combined into \textit{Others}. The band indicates the post-fit uncertainty. The lower panels show the ratio of data to the background expectation.  }
\label{fig:results:NN_control}
\end{figure}

Background-only likelihood fits
are performed for each NN. The obtained normalisation factors for the \ttbar, \Wjets, and single-top processes vary between the fits
for the different NNs: between $1.00\pm 0.28$ and $1.14 \pm 0.27$ for $t\bar{t}$, between $0.91\pm0.19$ and $1.08\pm0.17$ for \Wjets, and between $0.53\pm0.30$ and $0.60\pm0.23$ for single top.
The reduction of the single-top contribution appears large, but it is less than
the difference between the nominal and alternative schemes for modelling the interference
between the $t \bar{t}$ and $t W$ processes, which are described in Sections~\ref{sec:data_simulation} and~\ref{sec:syst_uncertainties}.
The \nnout distributions after the
fit are validated in the CRs by comparing data with simulation. As an example,
the plots for the NN training with a VLT signal with $\mathcal{B}(Zt,Ht,Wb) = (0.8,0.1,0.1)$ and a VLB signal with $\mathcal{B}(Zb,Hb,Wt) = (0.1,0.1,0.8)$
are shown in Figure~\ref{fig:results:NN_control}.
For the training with the VLT signal, the corresponding number of events expected from each process in the three CRs and also the
signal region is shown in Table~\ref{table:yields}, together with the number of observed events and the expected signal yield for a mass of \SI{1.2}{\TeV}.
The large uncertainty in the single-top yield due to the different schemes for modeling the interference can also be observed here.
The uncertainty in the total background is less than the uncertainty in the separate processes because of strong
(anti-)correlations between various systematic uncertainties.

\begin{figure}
\centering
\begin{subfigure}[b]{0.33\textwidth}
\centering
\includegraphics[width=\textwidth]{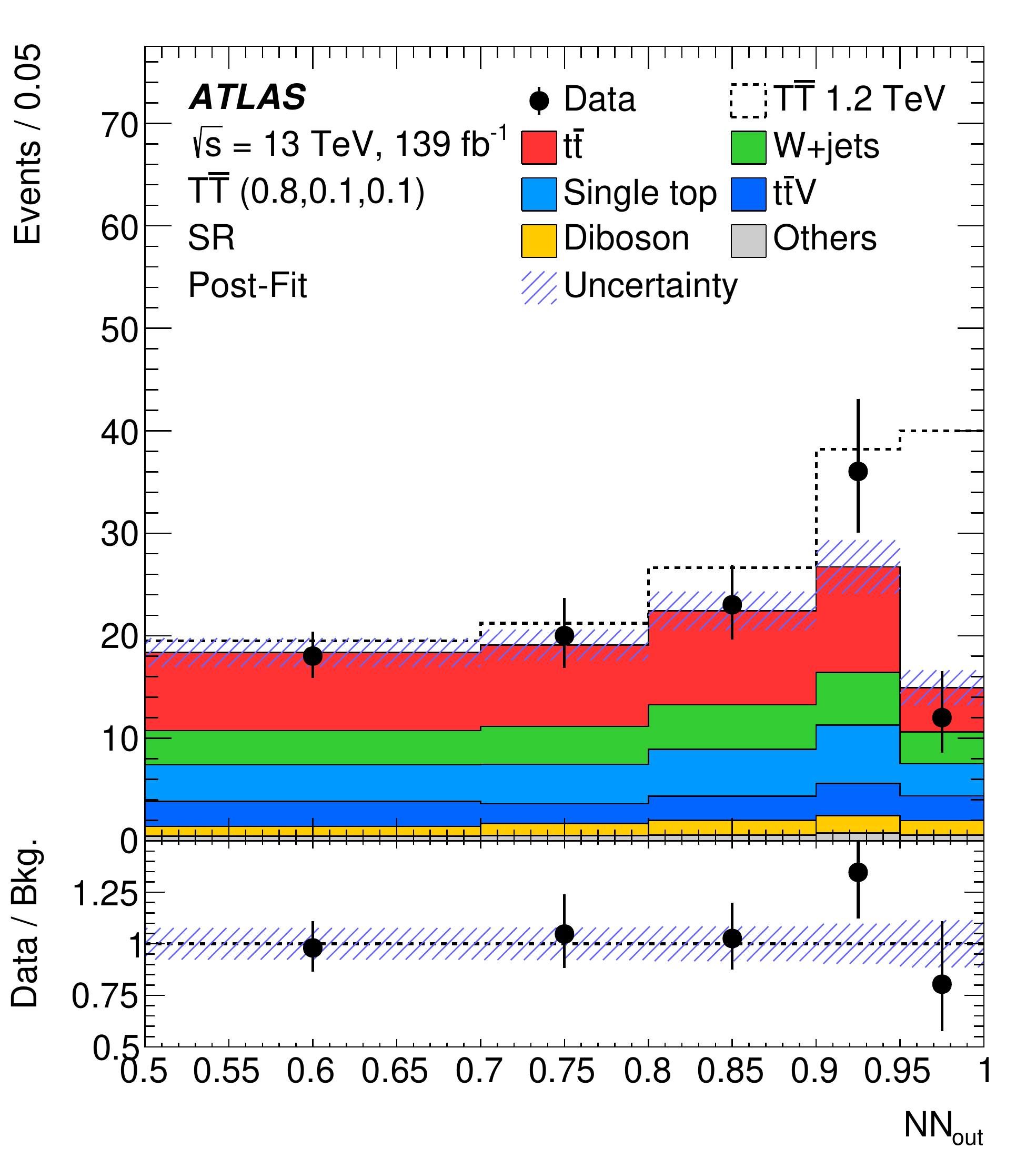}
\caption{}
\label{fig:results:postFit:SR_NN0}
\end{subfigure}\begin{subfigure}[b]{0.33\textwidth}
\centering
\includegraphics[width=\textwidth]{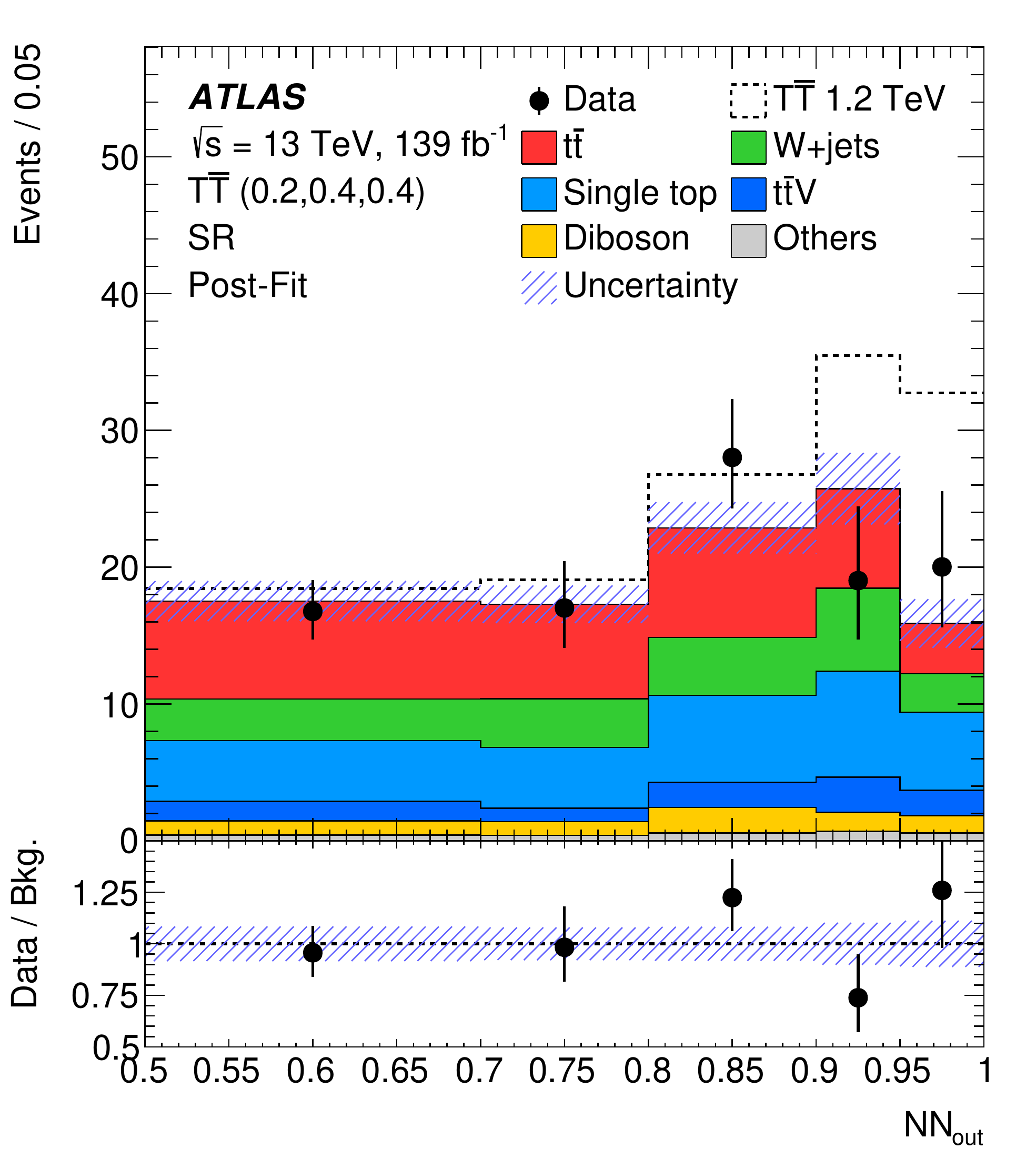}
\caption{}
\label{fig:results:postFit:SR_NN3}
\end{subfigure}\begin{subfigure}[b]{0.33\textwidth}
\centering
\includegraphics[width=\textwidth]{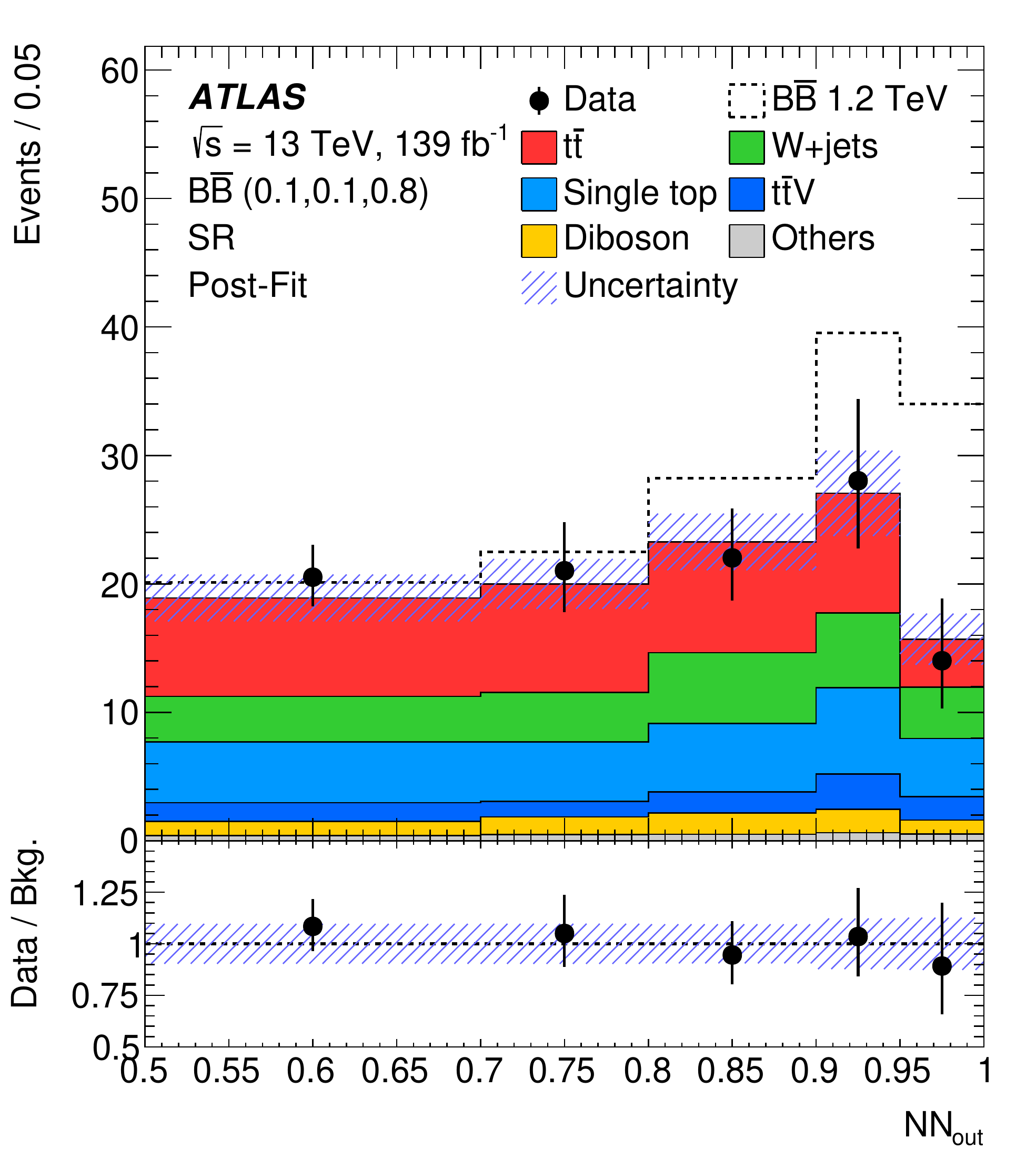}
\caption{}
\label{fig:results:postFit:SR_NN4}
\end{subfigure}
\caption{Data and background expectation in the signal region after the simultaneous background-only fit to data (Post-Fit) for a NN training for \subref{fig:results:postFit:SR_NN0} a VLT signal with $\mathcal{B}(Zt,Ht,Wb) = (0.8,0.1,0.1)$, \subref{fig:results:postFit:SR_NN3} a VLT signal with $\mathcal{B}(Zt,Ht,Wb) = (0.2,0.4,0.4)$, and \subref{fig:results:postFit:SR_NN4} a VLB signal with $\mathcal{B}(Zb,Hb,Wt) = (0.1,0.1,0.8)$. Contributions from \ttH, \tWZ, and \Zjets are combined into \textit{Others}. Expected pre-fit signal distributions with the signal branching ratio corresponding to the respective training are added on top of the background expectation, using a signal mass of \SI{1.2}{\TeV}. The band indicates the statistical and systematic uncertainties. The ratio of data to the background expectation is shown in the bottom panels of the plots.  }
\label{fig:results:postfit}
\end{figure}

The \nnout distributions in the signal region for the two training cases shown in Figure~\ref{fig:results:NN_control},
and for
another training for a VLT signal with $\mathcal{B}(Zt,Ht,Wb) = (0.2,0.4,0.4)$,
are shown in Figure~\ref{fig:results:postfit}.
No significant deviations from the SM expectation are observed for these three cases or when using other trained NNs.

\begin{table}[tb]
\centering
\caption{Observed data event yields and the expected background event yields with their total uncertainties in the control and signal regions after the background-only fit considering a NN trained for a VLT signal with branching ratio $\mathcal{B}(Zt,Ht,Wb) =(0.8,0.1,0.1)$. For comparison, the event yields for a VLT signal with a mass of \SI{1.2}{\TeV} and a branching ratio of $\mathcal{B}(Zt,Ht,Wb) = (0.8,0.1,0.1)$ are given. }
\begin{tabular}{lcccc}
\toprule
&     \Wjets CR     		&   Single-top CR   		&  Low-\nnout CR   		&       SR        \\ \midrule
$\ttbar $                                           		&  $1720 \pm 400$~~   	&  $520 \pm 130$   		& $730 \pm 140$  		& $79 \pm 31$   \\
\Wjets                                              		&  $2610 \pm 370$~~   	&  $256 \pm 47$~~    	& $231 \pm 54$~~   		& $37.7 \pm 9.1$~~  \\
Single top                                          		&  $170 \pm 150$   		&  $320 \pm 140$   		& $143 \pm 85$~~   		& $40 \pm 30$   \\
\ttV                                                		&  $20.4 \pm 3.2$~~   	&  $15.9 \pm 2.6$~~   	&  $132 \pm 21$~~   		& $23.9 \pm 4.0$~~  \\
Diboson                                             		&  $140 \pm 17$~~    	&  $14.2 \pm 2.4$~~   	& $65.4 \pm 8.9$~~  		& $12.1 \pm 1.8$~~  \\
\Zjets                                              		&  $14.0 \pm 3.8$~~   	&  $3.85 \pm 0.75$  		& $4.80 \pm 0.73$ 		& $1.14 \pm 0.13$ \\
\ttH                                                		&  $2.55 \pm 0.33$  		&  $5.59 \pm 0.69$  		& $10.9 \pm 1.3$~~  		& $2.27 \pm 0.33$ \\
\tWZ                                                		&  $0.90 \pm 0.12$  		&  $0.82 \pm 0.11$  		& $10.0 \pm 1.3$~~  		& $1.89 \pm 0.28$ \\ \midrule
Total background                                    	&  $4676 \pm 70$~~~~   	&  $1140 \pm 34$~~~~   	& $1322 \pm 37$~~~~  	& $198 \pm 14$~~   \\
Data                                                		&  4676        			&  1135        			& 1321      			&  206       \\ \midrule
$m_T = \SI{1.2}{\TeV}$                              &                   			&                   			&                 			&                 \\
$\mathcal{B}(Zt,Ht,Wb) = (0.8,0.1,0.1)$    & $0.35 \pm 0.12$ 		& $0.96 \pm 0.13$ 		& $7.01 \pm 0.51$ 		& $53.8\pm 1.8$~~ \\ \bottomrule
\end{tabular}
\label{table:yields}
\end{table}

Upper limits on the pair-production cross-sections for $T$ and $B$ quarks are calculated at the \SI{95}{\percent} CL. For each signal mass and branching ratio, the NN giving the most stringent expected limit is selected. These obtained cross-section limits are compared with the theoretical cross-section to set exclusion limits on the signal mass. The limits are calculated for $T$ and $B$ quarks in the weak-isospin singlet and doublet representations, with mass-dependent branching ratios, as well as for pure $T\rightarrow Zt$ and $B\rightarrow Wt$ decays, where the latter corresponds to
$X \rightarrow Wt$ as well as to the $(T,B)$ doublet as mentioned before.
For the doublet scenarios, the contribution from the VLQ partner is either not considered, leading to conservative limits,
or considered assuming mass-degenerate VLQs. Mass differences of at most a few \GeV\ are allowed, so that decays
from one member of the doublet to the other remain suppressed~\cite{Aguilar-Saavedra:2009xmz,Aguilar-Saavedra:2013qpa}.
Also for the mass-degenerate doublet scenario the seven NNs described in Section~\ref{nns} are used and the one with the most stringent expected limit is selected as described above, i.e., no additional NN is trained to potentially increase the sensitivity to the added yield from both doublet members.

\begin{figure}[ht]
\centering
\begin{subfigure}[b]{0.49\textwidth}
\centering
\includegraphics[width=0.9\textwidth]{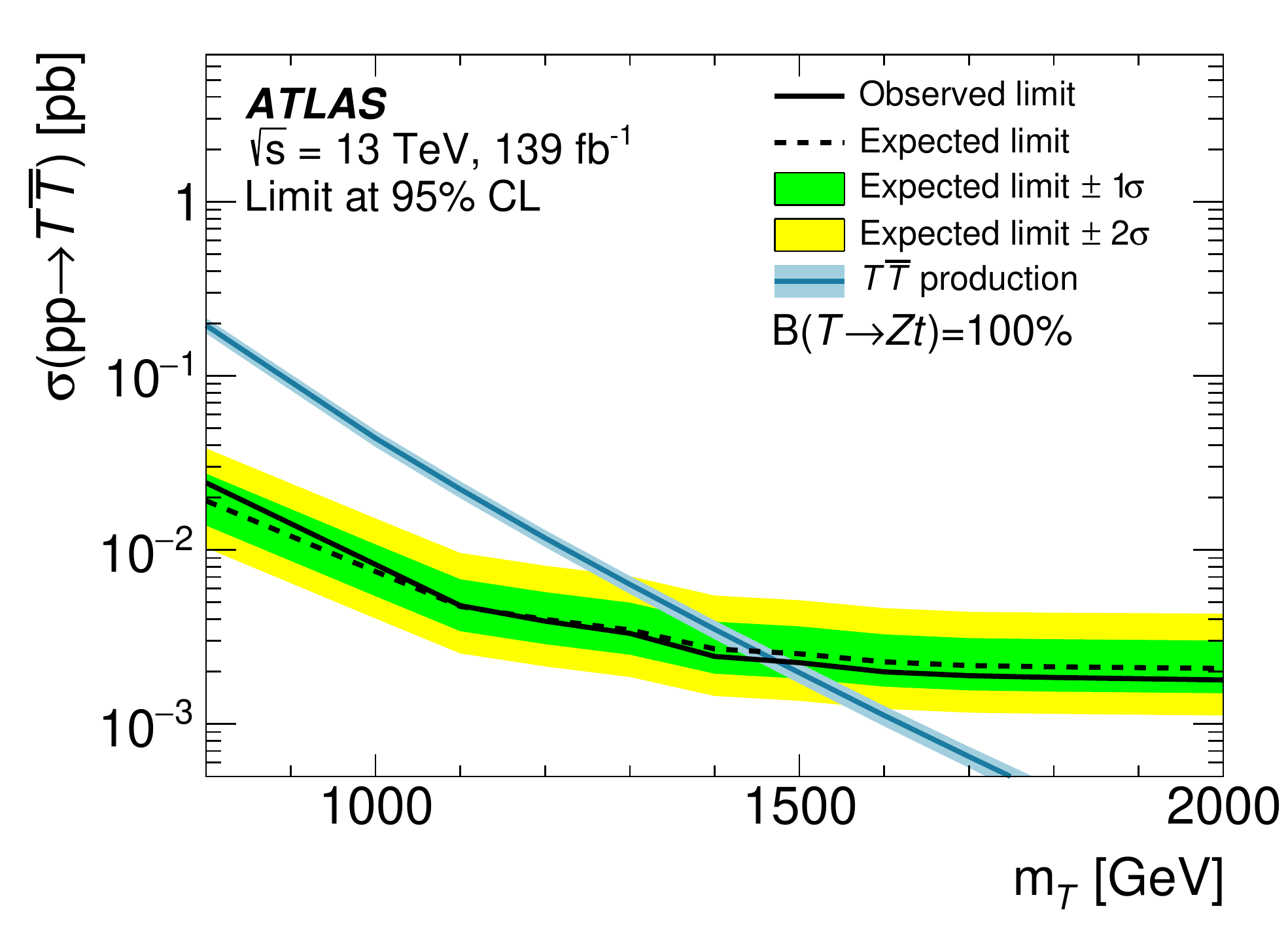}
\caption{}
\label{fig:results:1dlimit_Zt}
\end{subfigure}\begin{subfigure}[b]{0.49\textwidth}
\centering
\includegraphics[width=0.9\textwidth]{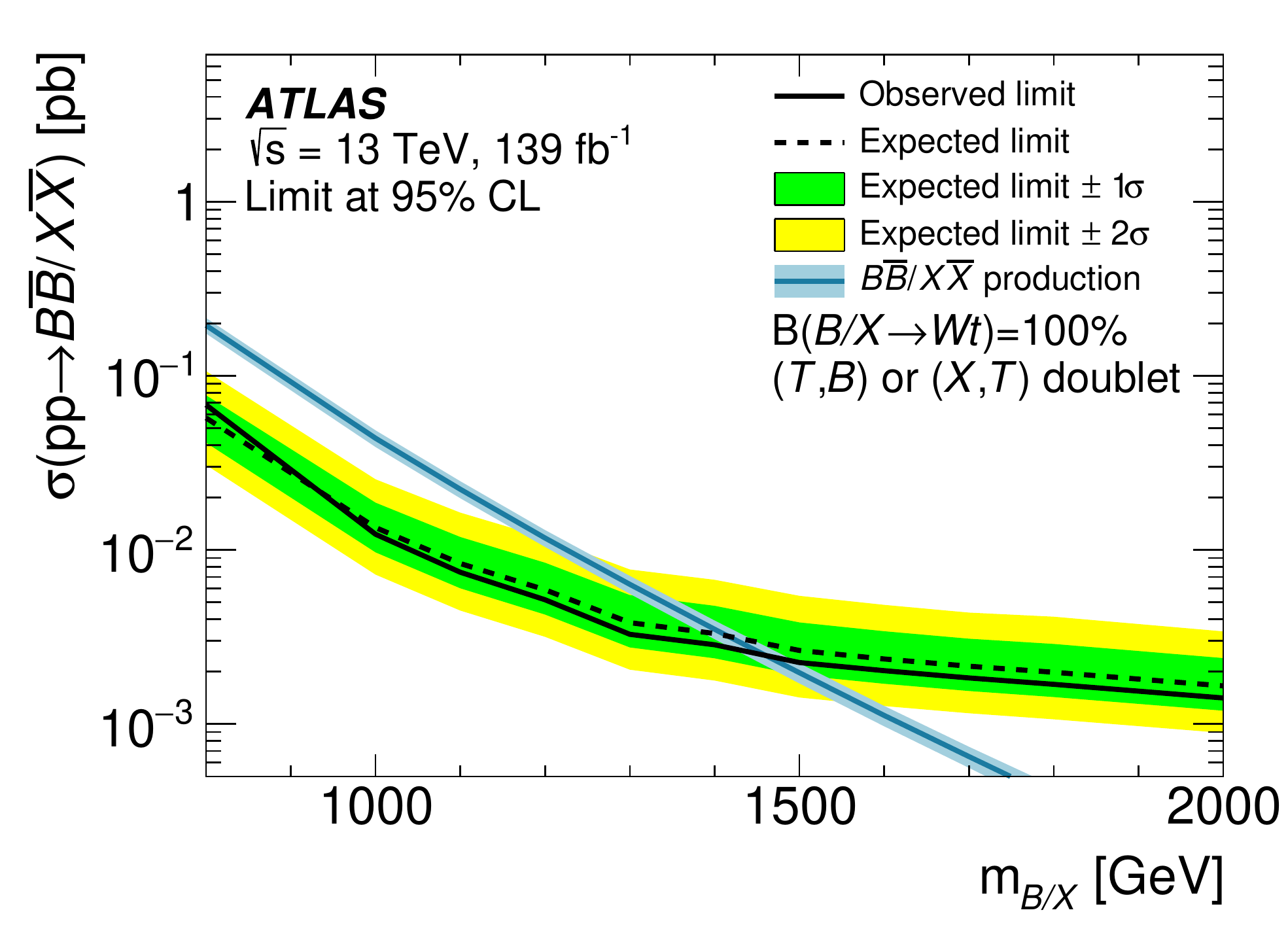}
\caption{}
\label{fig:results:1dlimit_Wt}
\end{subfigure}
 
\begin{subfigure}[b]{0.49\textwidth}
\centering
\includegraphics[width=0.9\textwidth]{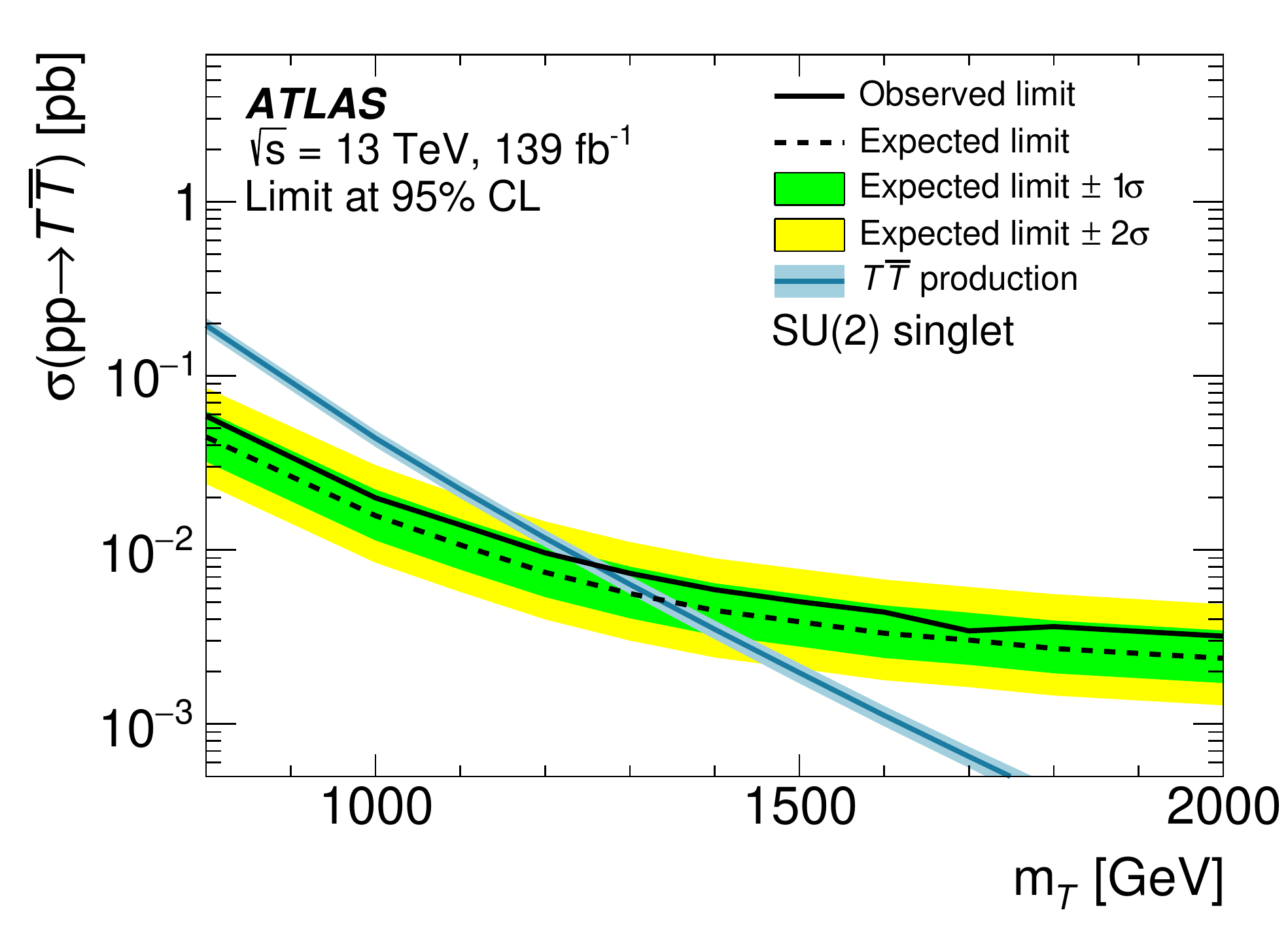}
\caption{}
\label{fig:results:1dlimit_T_singlet}
\end{subfigure}\begin{subfigure}[b]{0.49\textwidth}
\centering
\includegraphics[width=0.9\textwidth]{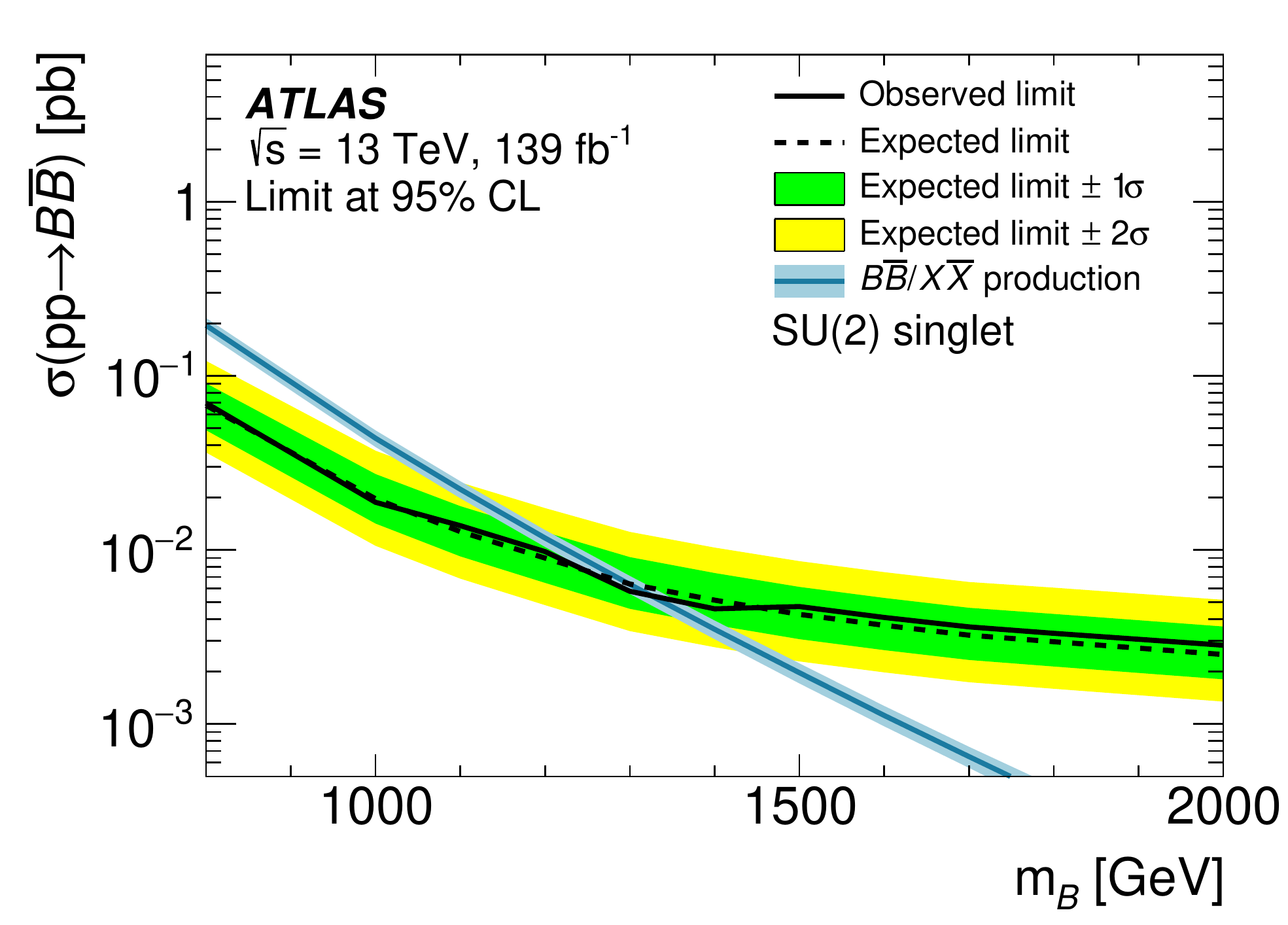}
\caption{}
\label{fig:results:1dlimit_B_singlet}
\end{subfigure}
 
\begin{subfigure}[b]{0.49\textwidth}
\centering
\includegraphics[width=0.9\textwidth]{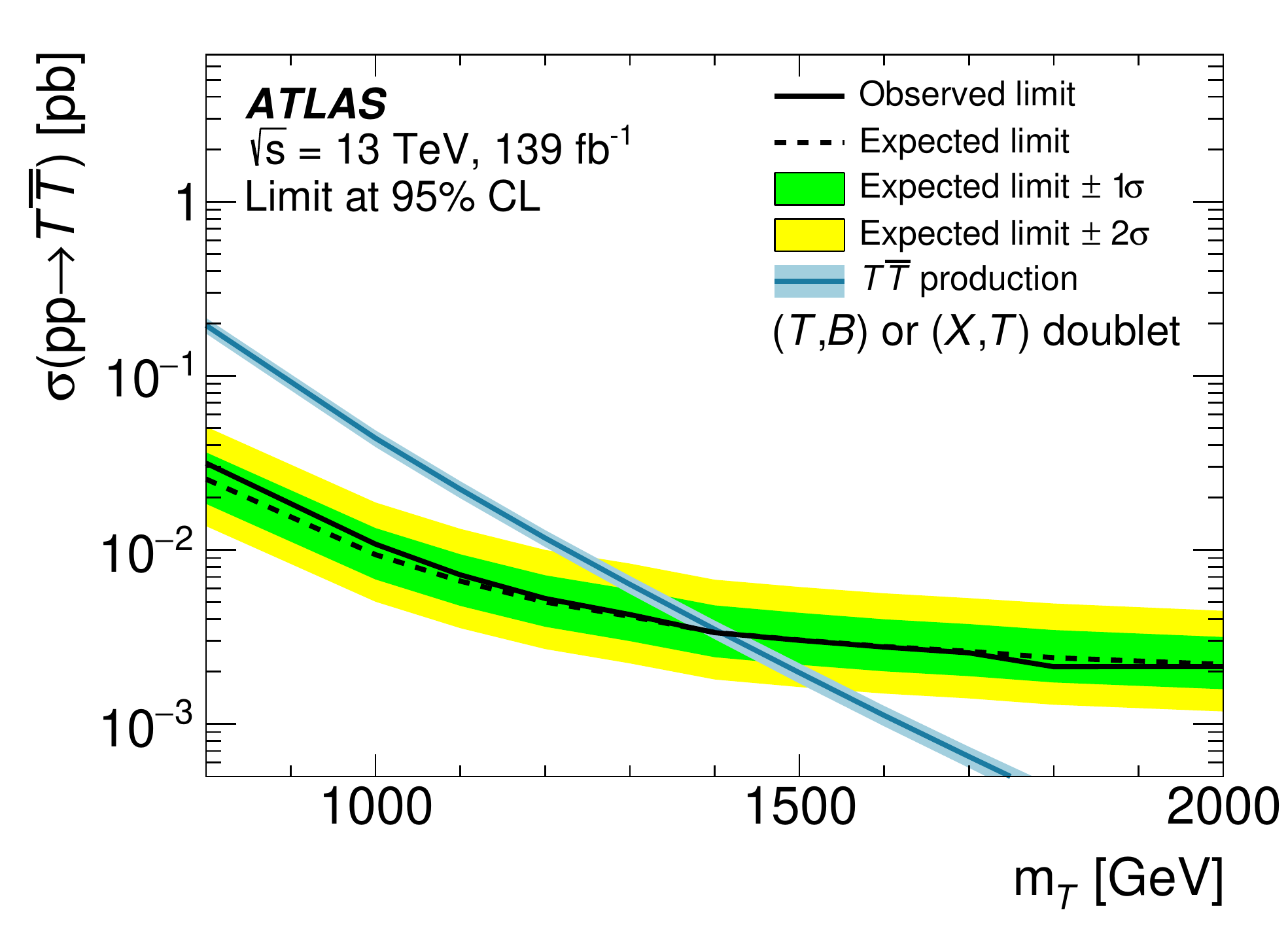}
\caption{}
\label{fig:results:1dlimit_T_doublet}
\end{subfigure}\begin{subfigure}[b]{0.49\textwidth}
\centering
\includegraphics[width=0.9\textwidth]{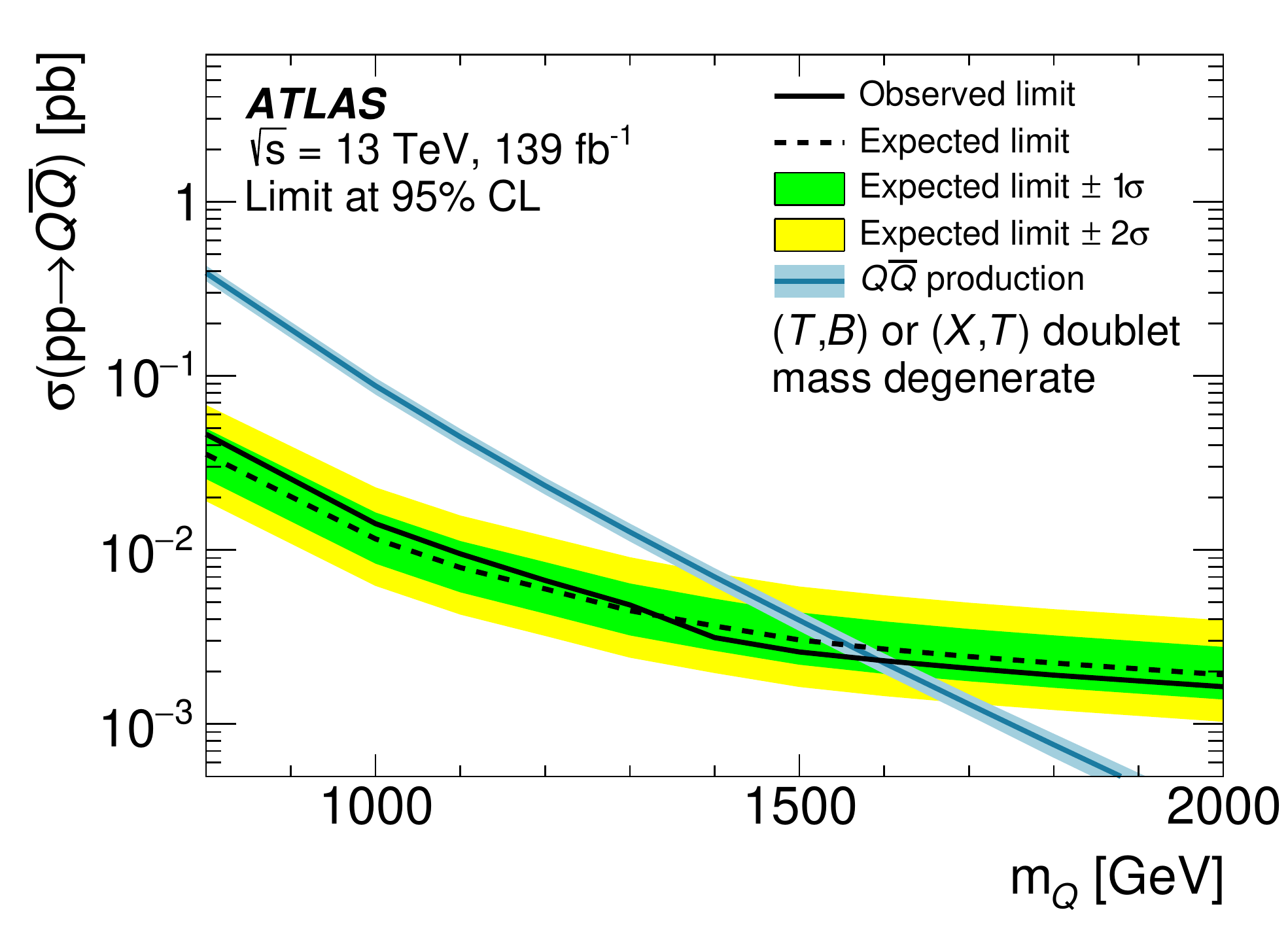}
\caption{}
\label{fig:results:1dlimit_TB_doublet}
\end{subfigure}
 
\caption{Expected and observed upper limits on the signal cross-section for \subref{fig:results:1dlimit_Zt} the case $\mathcal{B}(T\rightarrow Zt) = \SI{100}{\percent}$, \subref{fig:results:1dlimit_Wt} the case $\mathcal{B}(B\rightarrow Wt) = \SI{100}{\percent}$, \subref{fig:results:1dlimit_T_singlet} a $T$ quark in the SU(2) singlet representation, \subref{fig:results:1dlimit_B_singlet} a $B$ quark in the SU(2) singlet representation, and \subref{fig:results:1dlimit_T_doublet} a $T$ quark in an SU(2) doublet. In the doublet scenario, contributions from the not-considered vector-like quark are neglected, leading to conservative limits. The SU(2) $(T,B)$ doublet scenario considering contributions from both the $T$ and $B$ quark is shown in \subref{fig:results:1dlimit_TB_doublet}, assuming mass-degenerate VLQs. The thickness of the theory curve represents the theoretical uncertainty from the PDFs, scales, and strong coupling constant $\alphas$. }
\label{fig:limits_1D_singlet_doublet}
\end{figure}

The expected and observed lower limits on the VLQ mass in the aforementioned models are listed in Table~\ref{tab:expected_observed_limits} and
shown in Figure~\ref{fig:limits_1D_singlet_doublet}.
The impact of the statistical uncertainties on the mass limits is larger than that of the systematic uncertainties.
However, the latter is not negligible and for the case
of pure $T \rightarrow Zt$ ($B\rightarrow Wt$) decays it reduces the expected lower limit by about \SI{40}{\GeV} (\SI{70}{\GeV}) to
a value of \SI{1.45}{\TeV} (\SI{1.42}{\TeV}).
For the three $T$-quark scenarios in Table~\ref{tab:expected_observed_limits}, the obtained mass limits are
\SIrange{300}{400}{\GeV} higher than in the earlier ATLAS analysis in the same final state using a subset
of the \RunTwo data~\cite{EXOT-2016-15}.
This improvement is only partially due to the larger dataset, as the expected limits on the cross section
for a VLT mass of \SI{1.4}{\TeV} improved between a factor of 4.5 for the pure $T \rightarrow Zt$ and a factor
of 7.7 for the SU(2) singlet. Especially when the branching fraction into $Zt$ becomes smaller the major
effect stems from the training of neural networks at several branching ratios
instead of using a cut-and-count analysis with a single SR as done previously.
The obtained mass limits for the first five scenarios in Table~\ref{tab:expected_observed_limits} are also better than the corresponding
limits in the combination of all ATLAS results using \SI{36}{\ifb}~\cite{EXOT-2017-17}, apart from the $T$ singlet scenario where the
observed limit is weaker than the expected limit.
The strongest lower limits on the VLQ masses, \SI{1.59}{\TeV}, are derived for the
weak-isospin doublets
assuming mass-degenerate VLQs.

\begin{table}
\caption{Expected (Exp.) and observed (Obs.) mass limits for the pair production of specific VLQs ($T$, $B$, $X$) in certain decay scenarios
corresponding to SU(2) singlet or doublet representations or to the decay into just one specific final state.
In the doublet scenarios, contributions from the VLQ partner are not considered, leading to conservative limits, except for the last row
where the VLQs in the doublet are assumed to be mass degenerate. Since the analysis does not distinguish between particles and
antiparticles, the limits for $B\rightarrow Wt$ also apply to the vector-like $X$ because it decays exclusively into $Wt$. Similarly, the $(T, B)$ doublet scenarios correspond to $(X, T)$ doublet scenarios.}
\label{tab:expected_observed_limits}
\centering
\begin{tabular}{llcc}
\toprule
VLQ              & Scenario                                                                            & Exp.\ limit [\si{\TeV}] & Obs.\ limit [\si{\TeV}] \\ \midrule
$T$                & $\mathcal{B}(T\rightarrow Zt) = \SI{100}{\percent}$                                 & 1.45                  & 1.47                  \\
$T$                & singlet                                                                             & 1.33                  & 1.26                  \\
$T$                & $(T, B)$ or $(X, T)$ doublet                                                        & 1.41                  & 1.41                  \\
$B$                & singlet                                                                             & 1.30                  & 1.33                 \\
$B/X$              & $\mathcal{B}(B/X \rightarrow Wt) = \SI{100}{\percent}$ or $(T, B)$/$(X, T)$ doublet & 1.42                  & 1.46                  \\
$T/B/X$            & $(T, B)$ or $(X, T)$ doublet, mass degenerate                                       & 1.56                  & 1.59                  \\ \bottomrule
\end{tabular}
\end{table}
 
\begin{figure}[ht]
\centering
\begin{subfigure}[b]{0.49\textwidth}
\centering
\includegraphics[width=0.9\textwidth]{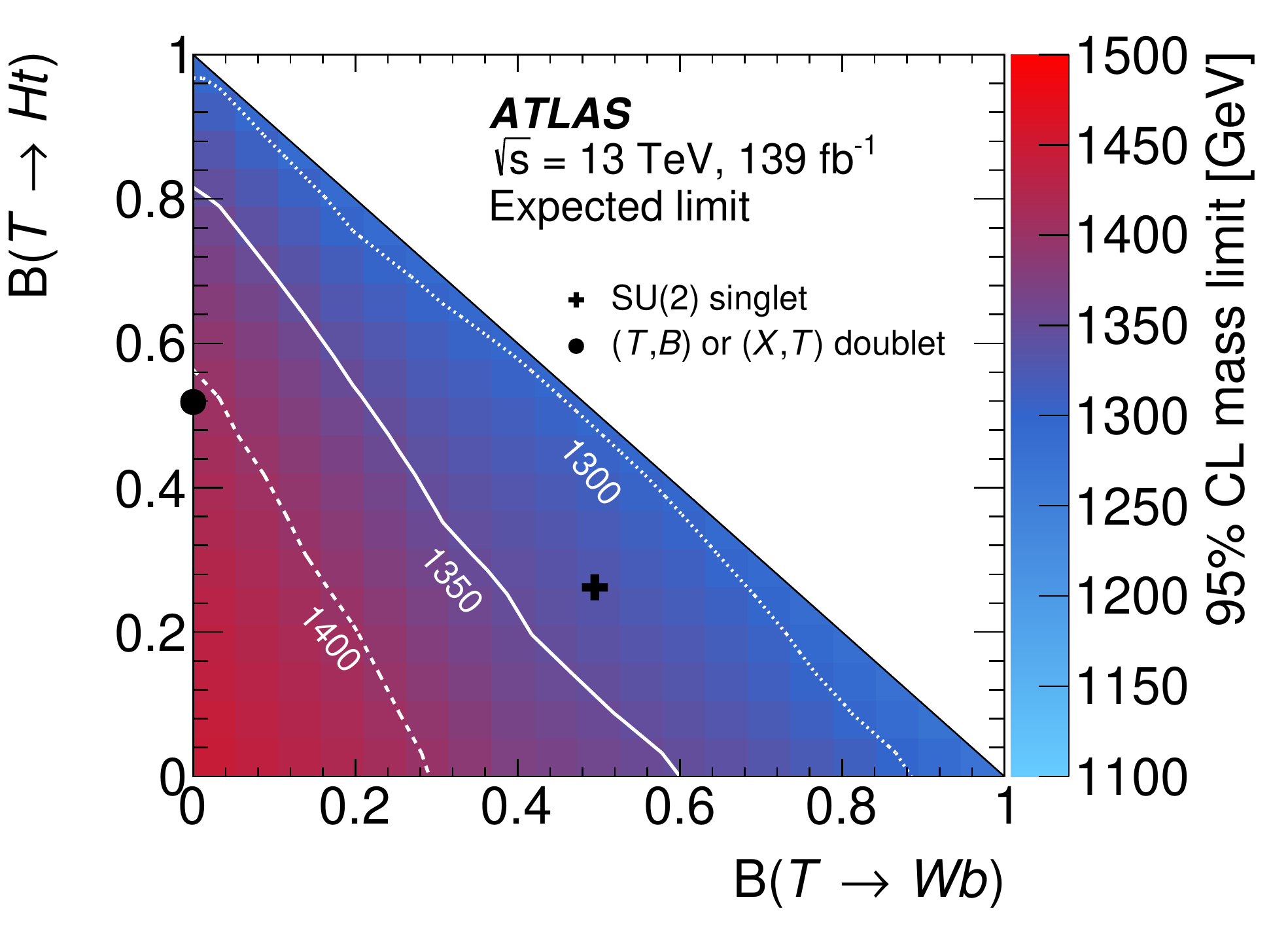}
\caption{}
\label{fig:results:2dlimit_VLT_exp}
\end{subfigure}\begin{subfigure}[b]{0.49\textwidth}
\centering
\includegraphics[width=0.9\textwidth]{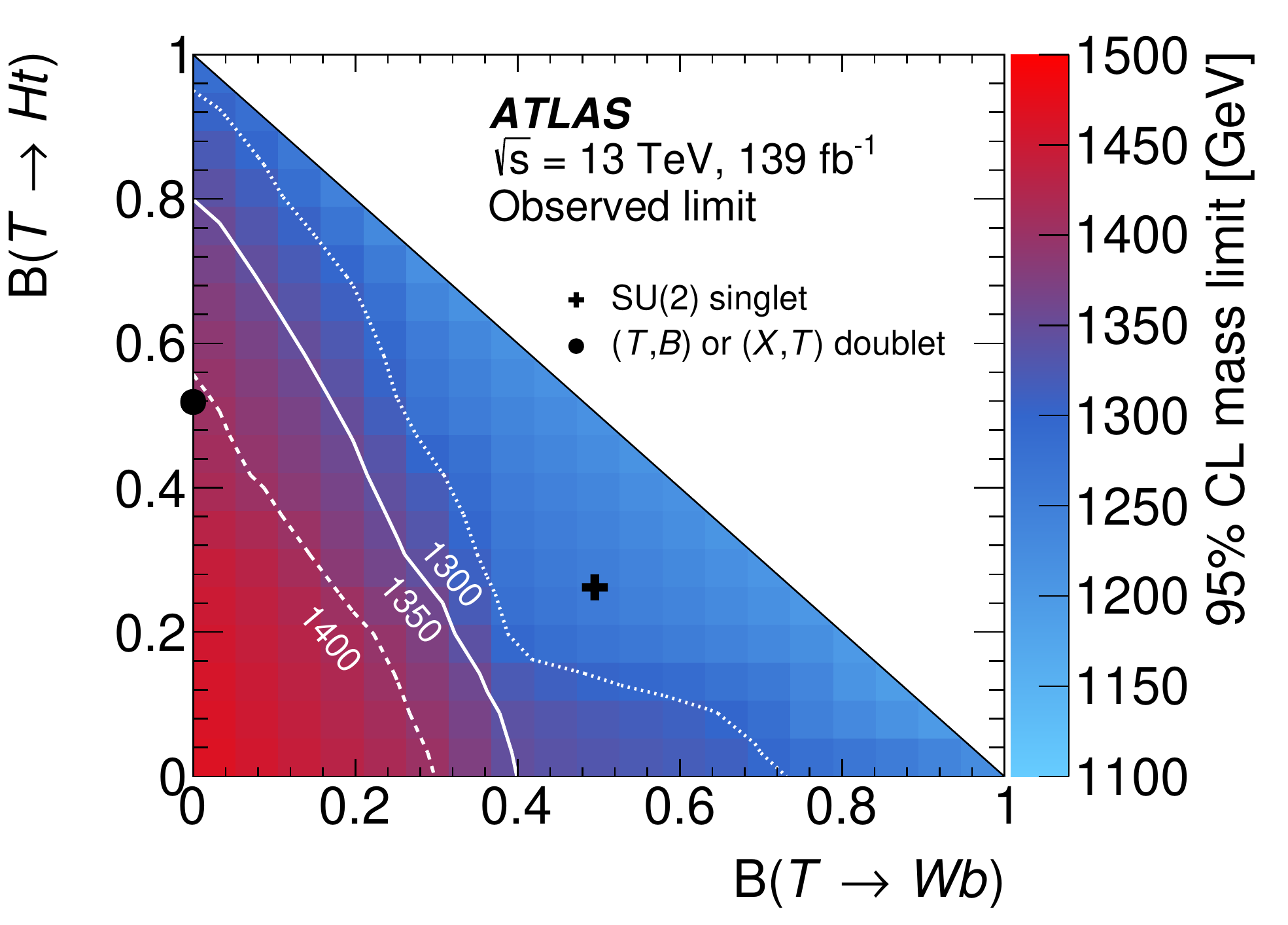}
\caption{}
\label{fig:results:2dlimit_VLT_obs}
\end{subfigure}
 
\begin{subfigure}[b]{0.49\textwidth}
\centering
\includegraphics[width=0.9\textwidth]{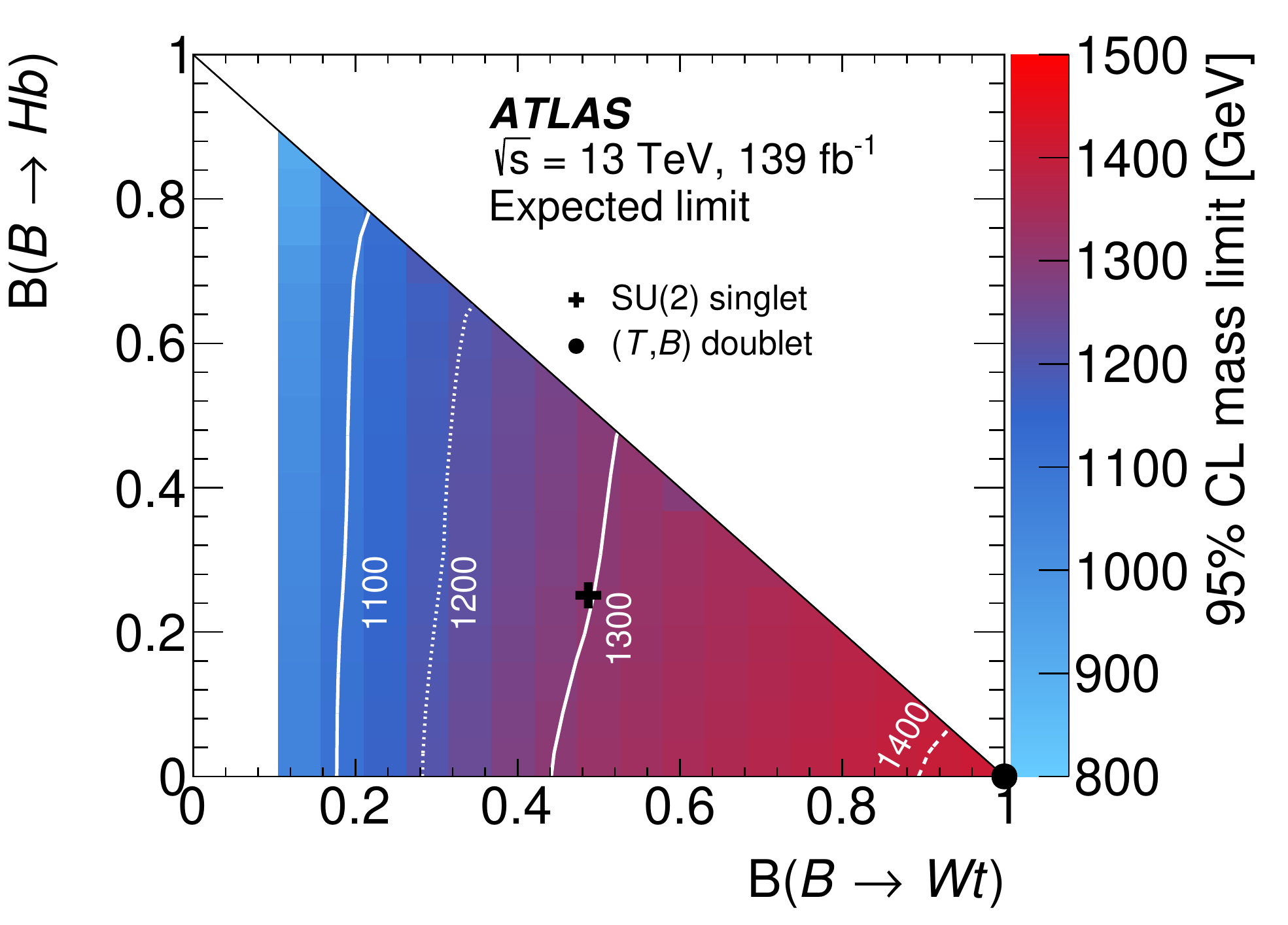}
\caption{}
\label{fig:results:2dlimit_VLB_exp}
\end{subfigure}\begin{subfigure}[b]{0.49\textwidth}
\centering
\includegraphics[width=0.9\textwidth]{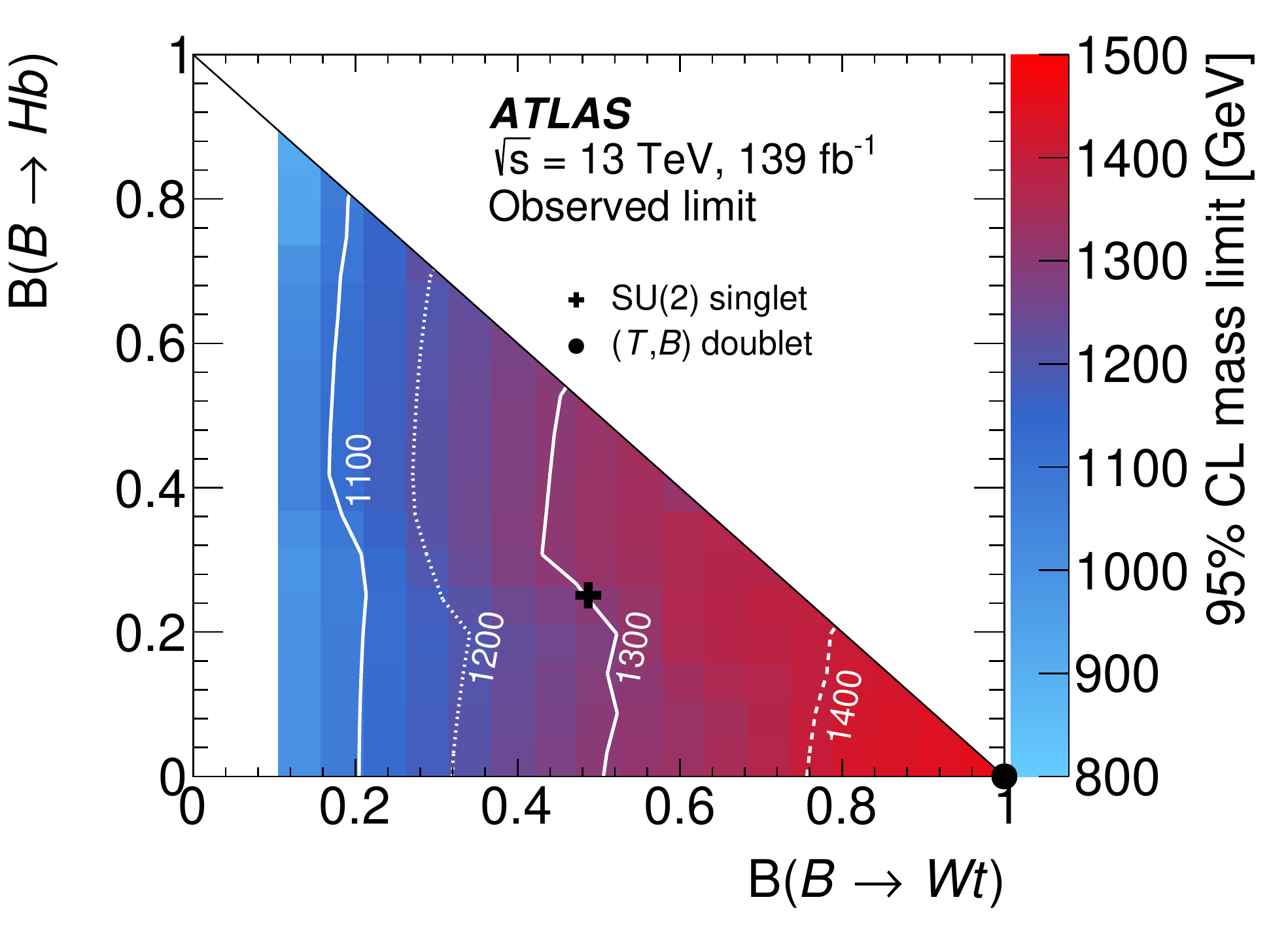}
\caption{}
\label{fig:results:2dlimit_VLB_obs}
\end{subfigure}
 
\caption{Expected (left) and observed (right) mass limits for $T\bar{T}$ (upper row) and $B\bar{B}$ (lower row) production. The mass limit is calculated using the NN giving the most stringent expected limit at each signal mass and branching ratio point. The white lines indicate mass exclusion contours. The black markers indicate the branching ratios for the SU(2) singlet and doublet scenarios for masses above \SI{800}{\GeV}, where they are approximately independent of the VLQ mass.
Since the analysis does not distinguish between particles and
antiparticles, the mass exclusion for the $B$ quark in the $(T,B)$ doublet is equivalent to the exclusion for the $X$ quark in the $(X,T)$ doublet. The white areas indicate that the mass limit is
below \SI{800}{\GeV}.
}
\label{fig:limits_2D_vlb}
\end{figure}

Apart from limits for specific models and branching ratios, lower limits on the signal mass are set as a function of the $T$ and $B$ branching ratios. The resulting expected and observed mass limits are shown in Figure~\ref{fig:limits_2D_vlb}. As expected, the highest sensitivity is found in the regions near $\mathcal{B}(T\rightarrow Zt)=\SI{100}{\percent}$ and $\mathcal{B}(B\rightarrow Wt)=\SI{100}{\percent}$. For the $T$ quark, the sensitivity for the mixed $ZtHt$ decay mode is larger than for the $ZtWb$ decay. In the case of the $B$ quark, the sensitivity decreases if the branching fraction for $B$ decay into a Higgs or $Z$ boson and a bottom quark increases.
The differences between the observed and expected limits for a vector-like $T$ quark around the SU(2) singlet branching ratio are not significant, as can be seen
in Figure~\ref{fig:results:1dlimit_T_singlet}. They result from the \nnout distribution obtained from the NN trained for a branching ratio $\mathcal{B}(Zt,Ht,Wb) = (0.2,0.4,0.4)$. In the last bin of the corresponding signal region in Figure~\ref{fig:results:postFit:SR_NN3}, the data slightly exceeds the predicted SM background.

\FloatBarrier
 
\section{Conclusion}
\label{sec:conclusion}

 
A search for pair-produced vector-like $T$ and $B$ quarks, with electric charge $+2/3$ and $-1/3$, respectively, is performed in events with exactly one isolated lepton, at least four jets including one that is $b$-tagged, and high missing transverse momentum. The analysis is based on data collected by the ATLAS experiment in $\sqrt{s}=13\,\TeV$ proton--proton collisions at the LHC, corresponding to an integrated luminosity of \SI{139}{\mbox{fb\(^{-1}\)}}. Several neural networks are trained for various branching ratios of the $T$ and $B$ quarks, assuming decays into a $W$, $Z$, or Higgs boson and a third-generation quark. The analysis considers all possible decays of the vector-like quarks, but it is most sensitive to the $T\rightarrow Zt$ and $B\rightarrow Wt$ decay modes.
Since the analysis does not distinguish between particles and antiparticles, the limits for $B\rightarrow Wt$ also apply
to a vector-like $X$ with electric charge $+5/3$.

No significant deviations from the Standard Model expectation are observed, and \SI{95}{\percent} CL upper limits on the pair-production
cross-sections for $T$ and $B$ quarks as a function of their mass are derived for various decay branching--ratio scenarios.
The lower limits on the masses of the $T$ and $B$ quarks in the weak-isospin singlet model are \SI{1.26}{\TeV} and \SI{1.33}{\TeV}, respectively, and \SI{1.41}{\TeV} for the $T$ quark in the doublet representation.
For the doublet, the contributions from the VLQ partner are not considered, leading to a conservative limit.
Stronger lower limits of \SI{1.47}{\TeV} and \SI{1.46}{\TeV} are set on the masses when considering pure $T \rightarrow Zt$ and $B\rightarrow Wt$ decays, where the latter corresponds to the $(T,B)$ or $(X,T)$ doublet and also applies to $X \rightarrow Wt$ decays.
For the three discussed $T$-quark scenarios, the obtained mass limits are \SIrange{300}{400}{\GeV} higher
than in the earlier ATLAS analysis in the same final state using a subset of the \RunTwo data.
The strongest lower limits for $T$, $B$ and $X$ are at \SI{1.59}{\TeV} for $(T,B)$ and $(X,T)$ weak-isospin doublets
where both VLQ partners are considered and assumed to be mass degenerate.
Finally, lower limits on the $T$ and $B$ quark masses are derived for all possible branching ratios.


\section*{Acknowledgements}


We thank CERN for the very successful operation of the LHC, as well as the
support staff from our institutions without whom ATLAS could not be
operated efficiently.
 
We acknowledge the support of
ANPCyT, Argentina;
YerPhI, Armenia;
ARC, Australia;
BMWFW and FWF, Austria;
ANAS, Azerbaijan;
CNPq and FAPESP, Brazil;
NSERC, NRC and CFI, Canada;
CERN;
ANID, Chile;
CAS, MOST and NSFC, China;
Minciencias, Colombia;
MEYS CR, Czech Republic;
DNRF and DNSRC, Denmark;
IN2P3-CNRS and CEA-DRF/IRFU, France;
SRNSFG, Georgia;
BMBF, HGF and MPG, Germany;
GSRI, Greece;
RGC and Hong Kong SAR, China;
ISF and Benoziyo Center, Israel;
INFN, Italy;
MEXT and JSPS, Japan;
CNRST, Morocco;
NWO, Netherlands;
RCN, Norway;
MEiN, Poland;
FCT, Portugal;
MNE/IFA, Romania;
MESTD, Serbia;
MSSR, Slovakia;
ARRS and MIZ\v{S}, Slovenia;
DSI/NRF, South Africa;
MICINN, Spain;
SRC and Wallenberg Foundation, Sweden;
SERI, SNSF and Cantons of Bern and Geneva, Switzerland;
MOST, Taiwan;
TENMAK, T\"urkiye;
STFC, United Kingdom;
DOE and NSF, United States of America.
In addition, individual groups and members have received support from
BCKDF, CANARIE, Compute Canada and CRC, Canada;
PRIMUS 21/SCI/017 and UNCE SCI/013, Czech Republic;
COST, ERC, ERDF, Horizon 2020 and Marie Sk{\l}odowska-Curie Actions, European Union;
Investissements d'Avenir Labex, Investissements d'Avenir Idex and ANR, France;
DFG and AvH Foundation, Germany;
Herakleitos, Thales and Aristeia programmes co-financed by EU-ESF and the Greek NSRF, Greece;
BSF-NSF and MINERVA, Israel;
Norwegian Financial Mechanism 2014-2021, Norway;
NCN and NAWA, Poland;
La Caixa Banking Foundation, CERCA Programme Generalitat de Catalunya and PROMETEO and GenT Programmes Generalitat Valenciana, Spain;
G\"{o}ran Gustafssons Stiftelse, Sweden;
The Royal Society and Leverhulme Trust, United Kingdom.
 
The crucial computing support from all WLCG partners is acknowledged gratefully, in particular from CERN, the ATLAS Tier-1 facilities at TRIUMF (Canada), NDGF (Denmark, Norway, Sweden), CC-IN2P3 (France), KIT/GridKA (Germany), INFN-CNAF (Italy), NL-T1 (Netherlands), PIC (Spain), ASGC (Taiwan), RAL (UK) and BNL (USA), the Tier-2 facilities worldwide and large non-WLCG resource providers. Major contributors of computing resources are listed in Ref.~\cite{ATL-SOFT-PUB-2021-003}.


\clearpage

\printbibliography
 
\clearpage
\input{atlas_authlist}

\end{document}

%% file: atlas_authlist.tex
 
\begin{flushleft}
\hypersetup{urlcolor=black}
{\Large The ATLAS Collaboration}

\bigskip

\AtlasOrcid[0000-0002-6665-4934]{G.~Aad}$^\textrm{\scriptsize 101}$,
\AtlasOrcid[0000-0002-5888-2734]{B.~Abbott}$^\textrm{\scriptsize 119}$,
\AtlasOrcid[0000-0002-7248-3203]{D.C.~Abbott}$^\textrm{\scriptsize 102}$,
\AtlasOrcid[0000-0002-1002-1652]{K.~Abeling}$^\textrm{\scriptsize 55}$,
\AtlasOrcid[0000-0002-8496-9294]{S.H.~Abidi}$^\textrm{\scriptsize 29}$,
\AtlasOrcid[0000-0002-9987-2292]{A.~Aboulhorma}$^\textrm{\scriptsize 35e}$,
\AtlasOrcid[0000-0001-5329-6640]{H.~Abramowicz}$^\textrm{\scriptsize 150}$,
\AtlasOrcid[0000-0002-1599-2896]{H.~Abreu}$^\textrm{\scriptsize 149}$,
\AtlasOrcid[0000-0003-0403-3697]{Y.~Abulaiti}$^\textrm{\scriptsize 116}$,
\AtlasOrcid[0000-0003-0762-7204]{A.C.~Abusleme~Hoffman}$^\textrm{\scriptsize 136a}$
% ...
\bigskip
\\

% ...
$^{29}$Physics Department, Brookhaven National Laboratory, Upton NY; United States of America.\\
% ...
$^{35}$$^{(a)}$Facult\'e des Sciences Ain Chock, R\'eseau Universitaire de Physique des Hautes Energies - Universit\'e Hassan II, Casablanca;$^{(b)}$Facult\'{e} des Sciences, Universit\'{e} Ibn-Tofail, K\'{e}nitra;$^{(c)}$Facult\'e des Sciences Semlalia, Universit\'e Cadi Ayyad, LPHEA-Marrakech;$^{(d)}$LPMR, Facult\'e des Sciences, Universit\'e Mohamed Premier, Oujda;$^{(e)}$Facult\'e des sciences, Universit\'e Mohammed V, Rabat;$^{(f)}$Institute of Applied Physics, Mohammed VI Polytechnic University, Ben Guerir; Morocco.\\
% ...
$^{55}$II. Physikalisches Institut, Georg-August-Universit\"{a}t G\"ottingen, G\"ottingen; Germany.\\
% ...
$^{101}$CPPM, Aix-Marseille Universit\'e, CNRS/IN2P3, Marseille; France.\\
$^{102}$Department of Physics, University of Massachusetts, Amherst MA; United States of America.\\
% ...
$^{116}$Department of Physics, New York University, New York NY; United States of America.\\
% ...
$^{119}$Homer L. Dodge Department of Physics and Astronomy, University of Oklahoma, Norman OK; United States of America.\\
% ...
$^{136}$$^{(a)}$Departamento de F\'isica, Pontificia Universidad Cat\'olica de Chile, Santiago;$^{(b)}$Millennium Institute for Subatomic physics at high energy frontier (SAPHIR), Santiago;$^{(c)}$Instituto de Investigaci\'on Multidisciplinario en Ciencia y Tecnolog\'ia, y Departamento de F\'isica, Universidad de La Serena;$^{(d)}$Universidad Andres Bello, Department of Physics, Santiago;$^{(e)}$Instituto de Alta Investigaci\'on, Universidad de Tarapac\'a, Arica;$^{(f)}$Departamento de F\'isica, Universidad T\'ecnica Federico Santa Mar\'ia, Valpara\'iso; Chile.\\
% ...
$^{149}$Department of Physics, Technion, Israel Institute of Technology, Haifa; Israel.\\
$^{150}$Raymond and Beverly Sackler School of Physics and Astronomy, Tel Aviv University, Tel Aviv; Israel.\\
% ...
$^{a}$ Also Affiliated with an institute covered by a cooperation agreement with CERN.\\
$^{b}$ Also at An-Najah National University, Nablus; Palestine.\\
$^{c}$ Also at Borough of Manhattan Community College, City University of New York, New York NY; United States of America.\\
$^{d}$ Also at Bruno Kessler Foundation, Trento; Italy.\\
$^{e}$ Also at Center for High Energy Physics, Peking University; China.\\
$^{f}$ Also at Center for Interdisciplinary Research and Innovation (CIRI-AUTH), Thessaloniki ; Greece.\\
% ...

\end{flushleft}
